\documentclass{article}
\usepackage{PRIMEarxiv}
\usepackage[utf8]{inputenc} 
\usepackage[T1]{fontenc}    
\usepackage{pznotations}
\usepackage{url}            
\usepackage{booktabs}      
\usepackage{amsfonts}      
\usepackage{nicefrac}  
\usepackage{microtype}   
\usepackage{lipsum}
\usepackage{fancyhdr}       
\usepackage{graphicx}       
\usepackage{mathtools}
\usepackage{subfig}
\graphicspath{{media/}}
\usepackage[flushleft]{threeparttable}
\usepackage{listings}
\usepackage{xcolor}
\usepackage[colorlinks]{hyperref}
\usepackage{cleveref}
\usepackage{enumitem}
\usepackage{multirow}
\usepackage{tikz}
\usepackage{wrapfig}
\usepackage{booktabs}
\usepackage{lipsum}
\usepackage{subfig}
\usepackage[numbers]{natbib}
\usepackage{amsthm}
\newtheorem{theorem}{Theorem}

\usepackage[final]{neurips_data_2024}
\newcommand{\algoname}{\texttt{LibMOON}}

\def \sixwidth{0.16}
\def \pentawidth{0.2}
\def \quadwid{0.24}

\lstdefinestyle{pythonstyle}{
    language=Python,
    basicstyle=\ttfamily\small,
    keywordstyle=\bfseries\color{blue},
    stringstyle=\color{red},
    commentstyle=\color{green}\itshape,
    numberstyle=\tiny\color{gray},
    numbers=left,
    numbersep=5pt,
    showspaces=false,
    showstringspaces=false,
    breaklines=true,
    frame=single,
    backgroundcolor=\color{yellow!10},
    captionpos=b
}

\title{LibMOON: A Gradient-based MultiObjective OptimizatioN Library in PyTorch}

\author{
    Xiaoyuan Zhang$^\clubsuit$, Liang Zhao$^\clubsuit$, Yingying Yu$^\clubsuit$, Xi Lin$^\clubsuit$, Yifan Chen$^\spadesuit$, \\
    \textbf{Han Zhao}$^\heartsuit$, \textbf{Qingfu Zhang}$^\clubsuit\thanks{Corresponding to Prof. Qingfu Zhang. Contact: \{xzhang2523-c@my.,qingfu.zhang@\}cityu.edu.hk. The first three authors contribute equally. Source code: \url{https://github.com/xzhang2523/libmoon}. Installment: \texttt{pip install libmoon}.
    } $ \\
    $^\clubsuit$ CityUHK,
    $^\spadesuit$ HKBU, $^\heartsuit$ UIUC.
}

\begin{document}
\maketitle
\begin{abstract}
    Multiobjective optimization problems (MOPs) are prevalent in machine learning, with applications in multi-task learning, learning under fairness or robustness constraints, etc. Instead of reducing multiple objective functions into a scalar objective, MOPs aim to optimize for the so-called Pareto optimality or Pareto set learning, which involves optimizing more than one objective function simultaneously, over models with thousands / millions of parameters. Existing benchmark libraries for MOPs mainly focus on evolutionary algorithms, most of which are zeroth-order / meta-heuristic methods that do not effectively utilize higher-order information from objectives and cannot scale to large-scale models with thousands / millions of parameters. In light of the above gap, this paper introduces \algoname, the first multiobjective optimization library that supports state-of-the-art gradient-based methods, provides a fair benchmark, and is open-sourced for the community.
\end{abstract}


\section{Introduction}
MultiObjective Optimization problems (MOPs) are ubiquitous in machine learning. For instance, trustworthy machine learning includes algorithmic fairness problems balancing the fairness level and the accuracy level~\cite{zhao2022inherent,xian2023fair}; 
in robotics, it is necessary to balance several objectives (e.g., forward speed and energy consumption) ~\cite{brandao2019multi, xu2020prediction}; similarly, recommendation systems face potentially conflicting objectives, such as novelty, accuracy, and diversity~\cite{jannach2022multi,keat9791369,zaizi2023102233}.
For all the applications above, the underlying optimization problem involves an MOP with \( m \) objectives and can be (informally) defined as:
\begin{equation}
    \min_\vtheta \vL(\vtheta) = ( L_1(\vtheta), \ldots, L_m(\vtheta) ),
    \label{eqn:mop}
\end{equation}
where \( L_1(\vtheta), \ldots, L_m(\vtheta) \) denote \( m \) (potentially) conflicting objectives and we denote the size of the model parameter as \(N \coloneqq |\vtheta| \). Note that as informally defined above,~\Cref{eqn:mop} is a vector optimization problem that does not necessarily admit a total ordering.
For a non-trivial MOP, no single solution can attain the minimum of all objectives simultaneously. To compare two solutions for an MOP, we introduce the concepts of \emph{dominance} and \emph{Pareto optimality}~\cite{ehrgott2005multicriteria}. 
Dominance occurs when solution \( \vtheta^{(a)} \) satisfies \( L_i(\vtheta^{(a)}) \leq L_i(\vtheta^{(b)}) \) for all \( 1 \leq i \leq m \), with at least one strict inequality.
A solution is Pareto optimal if no other solution in the feasible region dominates it. The set of all Pareto optimal solutions is called the Pareto set (PS), and its image set is called the Pareto front (PF).

Over the last few decades, multiobjective evolutionary algorithms (MOEAs) emerged as a widely used methodology for addressing MOPs due to their population nature to approximate the PS. Several popular MOEA libraries have emerged, including PlatEMO (Matlab) ~\cite{tian2017platemo}, Pagmo (C++)~\cite{Biscani2020}, and Pymoo (Python)~\cite{blank2020generating}. Compared to MOEAs, gradient-based multiobjective optimization (MOO) methods are particularly suitable designed for large-scale machine learning tasks involving thousands to millions of neural network parameters. While, gradient based MOO methods can only find Pareto \emph{stationary} solutions—solutions that cannot be \emph{locally} improved, in practice Pareto stationary solutions approximate global Pareto solutions well.

With the growing needs of gradient-based MOO methods for neural networks~\cite{sener2018multi, lin2019pareto,xu2020prediction,peitz2018gradient}, there is a pressing need for the development of a standard library to benchmark related algorithms and problems. For this reason, we introduce \algoname, the first modern gradient-based MOO library supporting over twenty state-of-the-art (SOTA) methods. We summarize our contributions as:
\begin{enumerate}[itemsep=-0.2em, topsep=0.0em, leftmargin=1.0em]
    \item We propose the \textit{first} modern gradient-based MOO library, called \algoname. \algoname~is implemented in PyTorch\cite{imambi2021pytorch} and carefully designed to support GPU acceleration. \algoname~supports synthetic problems, real-world problems, and MO machine learning problems such as fair classification, MO classification, MO regression, MO distribution matching and etc. 
    \item \algoname~supports over twenty SOTA gradient-based MOO methods for constructing PS/PF, including those using \emph{finite} solutions to approximate the whole PS/PF~\cite{mahapatra2020multi,liu2021profiling}, Pareto set learning (PSL)~\cite{navon2020learning,lin2020controllable} aimed at approximating the \emph{entire} PS/PF with a single neural model, and MOBO methods that designed to avoid frequent function evaluations for  
    \item \algoname~has already been open-sourced at \href{https://github.com/xzhang2523/libmoon}{Github}~\footnote{\url{https://github.com/xzhang2523/libmoon}} with document at \href{https://libmoondocs.readthedocs.io/en/latest/}{LibMOON Docs} ~\footnote{\url{https://libmoondocs.readthedocs.io/en/latest/}}. Beyond examining its source code, \algoname~can be installed via \texttt{pip install libmoon} as an off-the-shelf gradient-based multiobjective tool for industrial usage. 
\end{enumerate}

\paragraph{Notation.} In this paper, bold letters represent vector (e.g., $\vlam$ denotes a preference vector), while non-bold letters represent scalars. $\vx_k$ denotes vector $\vx$ at $k$-th iteration and $x_k$ denotes the $k$-th entry of $\vx$. The preference vector $\vlam$ lies in the $m$-dim simplex ($\mDelta_m$), satisfying $\sum_{i=1}^m \lambda_i=1$ and $\lambda_i \geq 0$. The decision network parameter $\vtheta$ has a size of $n$.
For two $m$-D vectors $\vx^{(a)}$ and $\vx^{(b)}$, $\vx^{(a)} \preceq \vx^{(b)}$ means $\vx_i^{(a)} \leq \vx_i^{(b)}$ for all $i \in [m]$; $\vx^{(b)}$, $\vx^{(a)} \preceq_\mathrm{strict} \vx^{(b)}$ means that $\vx_i^{(a)} \leq \vx_i^{(b)}$ for all $i \in [m]$ and for at least one index $j$, $\vx_j^{(a)} < \vx_j^{(b)}$. $\vx^{(a)} \prec \vx^{(b)}$ means that $\vx_i^{(a)} < \vx_i^{(b)}$ for all $i \in [m]$. 
Refer to~\Cref{tab:notation} for the full notation meanings.

\section{Related works}
\subsection{Gradient-based multiobjective optimization} \label{sec:gbmoo}
Gradient-based MOO has a long-standing research history. For example, the well-known convex optimization \cite{boyd2004convex}[Chap 4] book outlines how linear scalarization can transform an MOO problem into a single-objective optimization (SOO) problem. However, for much of the past few decades, gradient-based methods have not been the primary approach for MOO, with MOEAs gaining more prominence due to their population-based approach, which is well-suited for approximating the PS and avoiding local optima. In recent years, however, gradient-based MOO has experienced a resurgence, particularly in (deep) machine learning, where these methods scale better with the number of decision variables. A pivotal contribution in this area is the MGDA-UB~\cite{sener2018multi}, which introduced MOO techniques into deep learning by casting multi-task learning (MTL) as an MOO problem. Since then, many approaches have followed, including EPO~\cite{mahapatra2020multi}, Pareto Multi-Task Learning (PMTL)~\cite{lin2019pareto}, MOO with Stein Variational Gradient Descent (MOO-SVGD)~\cite{liu2021profiling}, and methods for learning the entire PS~\cite{lin2020controllable,navon2020learning,zhong2024panacea,lin2024smooth,lin2022pslmobo}. To make a fair comparison and for the ease of developing new methods, this paper implements them following a standardized manner.

\subsection{Multiobjective optimization libraries} \label{sec:moolib}
A number of multiobjective libraries exist before our work. LibMTL~\cite{lin2023libmtl} is a library for multitask learning; the difference is that \algoname~is designed to study the distributions of Pareto solutions, while LibMTL is mainly designed to study how to find a single solution that benefits all objectives. The major difference between \algoname~and other previous libraries is \algoname~is a gradient-based library, while others are evolutionary computation-based libraries. Comparisons are provided in~\Cref{tab:library_comparison}.

\textbf{LibMTL}~\cite{lin2023libmtl} is a Python-based library for multitask learning. LibMTL aims to find a single network to benefit all tasks, such as calculating a benign updating direction or optimizing a network architecture. In contrast, \algoname~addresses inherent trade-offs in machine learning problems, where improving one objective inevitably worsens others, and explores the distribution of Pareto solutions.

\textbf{jMetal}~\cite{jMetal}, \textbf{Pymoo}~\cite{pymoo} and \textbf{PlatEMO}~\cite{tian2017platemo} are Java, python and Matlab frameworks for MOEAs, supporting popular methods such as NSGA-III~\cite{nsga31,nsga32}, MOEA/D~\cite{zhang2007moea}, and SMS-EMOA \cite{beume2007sms}. Pymoo allows flexible algorithm customization with user-defined operators and data visualization. PlatEMO is a MATLAB-based multiobjective optimization tool supporting over 160 MOEAs and a comprehensive test problems, including sparse, high-cost, large-scale, and multimodal. PlatEMO also contains a number of metrics and supporting visualization during the optimization process. 

\textbf{Pagmo}~\cite{Biscani2020} is a C++ library for parallel multiobjective global optimization, utilizing evolutionary algorithms and gradient-based methods like simplex, SQP, and interior-point techniques. It supports constrained, unconstrained, single- and multi-objective, continuous, integer, stochastic, and deterministic optimization problems.

\begin{wraptable}{r}{0.6\textwidth} 
\centering
\vspace{-15pt}
\setlength\tabcolsep{2.5 pt}
\caption{Previous MOO libraries and \algoname.}
\label{tab:library_comparison}
\tiny
    \begin{tabular}{lcclp{4cm}} 
    \toprule
    \textbf{Name} & \textbf{Language} & \textbf{Year} & \textbf{Key Features} \\ 
    \midrule
    \textbf{Pymoo}     & Python  & 2020 & \begin{tabular}[l]{@{}l@{}}(1) Evolutionary computation (EC) \\ (2) Zero-order methods \\ (3) Diverse problem types\end{tabular} \\
    \midrule
    \textbf{jMetal}    & Java    & 2011 & \begin{tabular}[l]{@{}l@{}}(1) Single-/multi-objective optimization \\ (2) Parallel algorithms \\ (3) Diverse problem types\end{tabular} \\
    \midrule 
    \textbf{PlatEMO}   & Matlab  & 2017 & \begin{tabular}[l]{@{}l@{}}(1) Over 160 MOEAs \\ (2) Various figure demonstrations \\ (3) Powerful and friendly GUI\end{tabular} \\ 
    \midrule 
    \textbf{Pagmo}     & C++     & 2020 & \begin{tabular}[l]{@{}l@{}}(1) Global optimization \\ (2) Parallel optimization\end{tabular} \\
    \midrule
    \textbf{LibMTL}    & Python  & 2023 & \begin{tabular}[l]{@{}l@{}}(1) Unified codebase \\ (2) Comprehensive SOTA MTL methods \\ (3) Flexible extension for new methods\end{tabular} \\
    \midrule
    \textbf{EvoTorch}  & Python  & N/A & \begin{tabular}[l]{@{}l@{}}(1) Distribution-based search algorithms \\ (2) Population-based search algorithms \\ (3) Multiple CPUs, GPUs, computers\end{tabular} \\
    \midrule
    \textbf{EvoX} & Python  & 2024 & \begin{tabular}[l]{@{}l@{}}(1) GPU acceleration optimization \\ (2) Single-/multi-objective optimization \\ (3) Neuroevolution/RL tasks\end{tabular}  \\
    \midrule
    \textbf{LibMOON}   & Python  & 2024  & \begin{tabular}[l]{@{}l@{}}(1) GPU-accelerated gradient solvers \\ (2) Pareto set learners \\ (3) Large-scale (millions \# params.) ML tasks\end{tabular} \\
    \bottomrule
    \end{tabular}
    \vspace{-55pt}
\end{wraptable}

\textbf{EvoTorch}~\cite{toklu2023evotorch} and \textbf{EvoX}~\cite{huang2024evox}.
EvoTorch accelerates evolutionary algorithms in PyTorch, while EvoX scales them to large, complex problems with GPU-accelerated parallel execution for single and multiobjective tasks, including synthetic problems and reinforcement learning.

\section{\algoname: A gradient-based MOO library in PyTorch} \label{sec:method}
This section introduces \algoname. We introduce its framework in \Cref{sec:framework}, and briefly introducing its supporting \textbf{problems} and \textbf{metrics}. Then we introduce supported \textbf{solvers} in~\Cref{sec:moo_solvers,sec:psl,sec:mobod}.

\begin{wrapfigure}{r}{0.6\textwidth}
    \centering
    \vspace{-15pt}
    \includegraphics[width=0.6\textwidth]{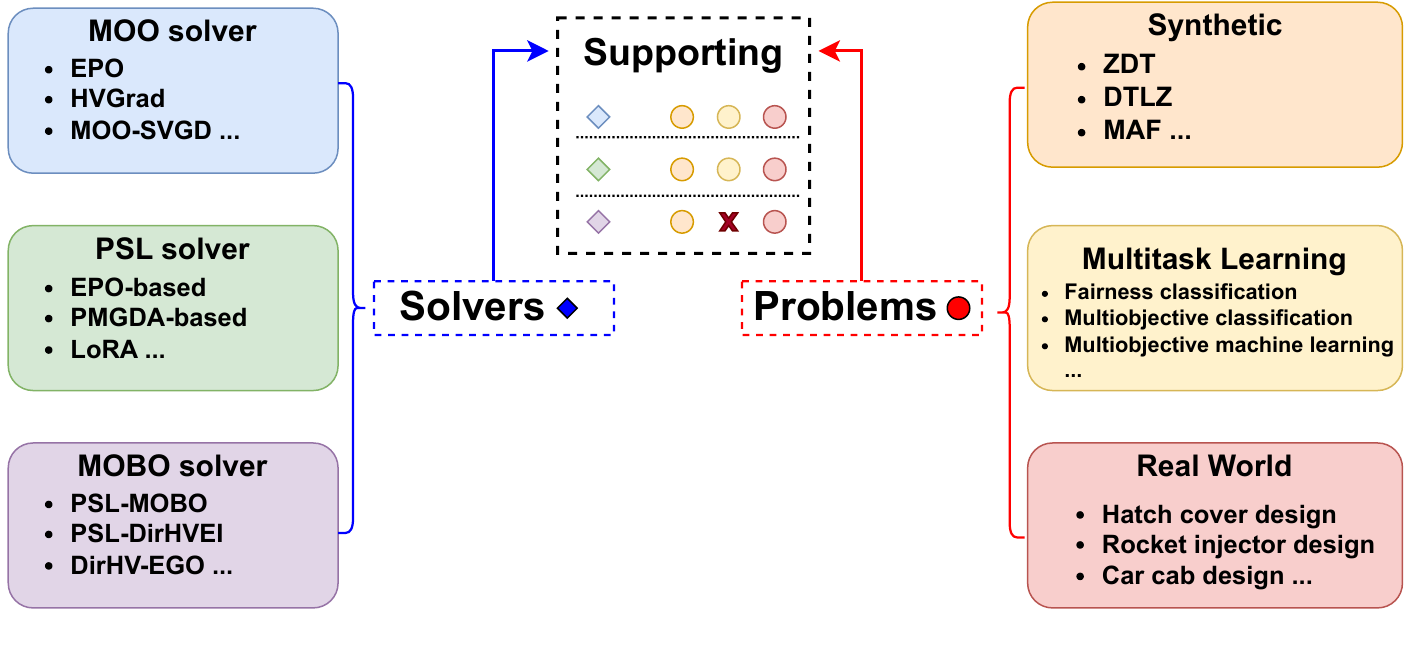}
    \caption{
    \textbf{Supported solvers and problems in \algoname}: \algoname~addresses synthetic, real-world and MTL problems with three categories of solvers: MOO, PSL and MOBO solvers. 
    }
    \vspace{-20pt}
    \label{fig:framework}
\end{wrapfigure}

\subsection{Framework} \label{sec:framework}
\Cref{fig:framework} demonstrates \algoname~components, including three categories of solvers: MOO solvers aiming to find a finite set of Pareto solutions satisfying certain requirements, Pareto set learning (PSL) solvers aiming to learn whole PS with a single model, and MOBO solvers aiming to solve expensive MO problems. Each solver category is designed highly modulized and new solvers are easy to plugin \algoname~by rewriting only a small portion of code, e.g., the specific algorithm of gradient manipulations~\footnote{An example of adding a new solver is provided in the \href{https://libmoondocs.readthedocs.io/en/latest/develop/add_method.html}{\algoname~Doc}.}. MOO and PSL solvers support all synthetic, MTL, and realworld (RE) problems, while MOBO solvers support synthetic and RE problems.

\paragraph{Supported problems.}
\algoname~currently supports three categories of methods, synthetic problems, MTL problems, and RE problems. 

\begin{wraptable}{r}{0.6\textwidth}
    \centering
    \vspace{-10pt}
    \caption{Supported MO machine learning problems.} \label{tab:supp:problems}
    \footnotesize
    \begin{threeparttable}
    \begin{tabular}{llllll}
        \toprule
        Method & $L_1$ & $L_2$ \\
        \midrule
        Fairness classification~\cite{ruchte2021scalable} & BCE & DEO \\
        MO classification~\cite{lin2019pareto} & CE - BR & CE - UL \\
        MO regression~\cite{hu2023revisiting} & MSE & MSE \\
        MO distribution matching & $D(\cdot \| \cdot)$ & $D(\cdot \| \cdot)$ \\
        \bottomrule
    \end{tabular}
    \begin{tablenotes}
        \item \noindent \tiny BCE: Binary Cross Entropy; DEO~\cite{padh2021addressing}: Difference of Equality of Opportunity; CE: Cross Entropy; BR: Bottom Right; UL: Upper Left; MSE: Mean Square Error. 
    \end{tablenotes}
    \end{threeparttable}
\end{wraptable}

\paragraph{Supported metrics.}
\algoname~supports a number of metrics, (1) hypervolume (HV), (2) inverted general distance (IGD), (3) fill distance (FD), (4) minimal distance ($\mathrm{l}_{\min}$), (5) smooth minimal distance ($\mathrm{sl}_{\min}$), (5) Spacing, (6) Span, (7) penalty-based intersection (PBI), (8) inner product (IP), (9) cross angle ($\vartheta$). Full descriptions of these indicators are provided in~\Cref{sec:metrics}.

\subsection{MOO solvers} \label{sec:moo_solvers}

In this paper, we specially mean MOO solvers are solvers to find a set of Pareto solutions. The most simple and commonly used method is convert the MOO problem to a single objective optimization problem through some aggregation functions. 
\paragraph{Aggregation-based methods.}
A straightforward way is to use some aggregation functions $g_\vlam(): \mathbb{R}^m \mapsto \mathbb{R}$ to convert a MOP to a single objective optimization problem. The reason that optimizing this converted single objective optimization problem will yield Pareto optimal solutions is due to the following two theorems.

\begin{theorem}[Adapted from Theorem 2.6.2~\cite{miettinen1999nonlinear}] \label{thm:po}
    If \( g_\vlam(\cdot) \) is \textbf{strictly decreasing} w.r.t vector \( \vL(\vtheta) \), i.e., \( g_\vlam(\vL(\vtheta^{(a)})) < g_\vlam(\vL(\vtheta^{(b)})) \) when \( \vL_i(\vtheta^{(a)}) \preceq_\mathrm{strict} \vL_i(\vtheta^{(b)}) \), then the optimal solution $\theta^*$ of \( g_\vlam(\vL(\vtheta)) \) serves as a \textbf{Pareto optimal} solution for the original MOP.
\end{theorem}
\textit{Proof.} See Mitten's book~\cite{miettinen1999nonlinear}, Page 22. Similarly, for only decreasing aggregation functions, we have the following theorem.
\begin{theorem} \label{thm:wpo}
    If \( g_\vlam() \) is \textbf{decreasing} w.r.t. vector $\vL(\vtheta)$ (i.e., $g_\vlam(\vL(\vtheta^{(a)})) \leq g_\vlam(\vL(\vtheta^{(b)}))$ when $\vL(\vtheta^{(a)}) \preceq_\mathrm{strict} \vL(\vtheta^{(b)})$, then the optimal solution $\theta^*$ of \( g_\vlam(\vL(\vtheta)) \) serves as a \textbf{weakly Pareto optimal} solution for the original MOP.
\end{theorem}
\textit{Proof.} Similar to the proof of Mitten's book~\cite{miettinen1999nonlinear}, Page 22. 

In \Cref{thm:wpo}, \emph{weakly Pareto optimality} means that for a solution $\vtheta^{(a)}$, no other solutions $\vtheta'$ can strictly dominate it, i.e., $\vL_i(\vtheta') < \vL_i(\vtheta^{(a)})$ for all $i \in [m]$.
Some common aggregation functions include the linear scalarization (LS) function, where \( g_\vlam^\mathrm{LS}(\vL(\vtheta)) = \sum_{i=1}^m \lambda_i L_i(\vtheta) \), the Tchebycheff function, where \( g_\vlam^\mathrm{Tche}(\vL(\vtheta)) = \max_{i \in [m]} \lambda_i \cdot (L_i(\vtheta) - z_i)  \) ($\vz$ is a reference point), Penalty-Based Intersection (PBI) function~\cite{zhang2007moea}, and COSMOS function~\cite{ruchte2021scalable}. For expressions and other aggregation functions, please refer to \Cref{sec:agg}. 

For a preference vector $\vlam \succ \bm 0$, $g_\vlam^\mathrm{LS}(\cdot)$ is a \emph{strict decreasing function} meaning that directly optimizing $g_\vlam^\mathrm{LS}(\cdot)$ yields \emph{Pareto optimal solutions} (by \Cref{thm:po}). However, for any $\vlam \in \mDelta_m$, optimizing $g_\vlam^\mathrm{Tche}(\cdot)$ only yields weakly \emph{Pareto optimal solutions} (by \Cref{thm:wpo}). For an improper setting of the weight factor $\mu$ (see \Cref{sec:agg} item 1 and 6), the optimal solution of $g_\vlam^\mathrm{PBI}(\cdot)$ or $g_\vlam^\mathrm{COSMOS}(\cdot)$ can be non-(weakly) Pareto optimal solutions of the original MOP. 

An aggregation function is optimized by gradient descent in \algoname~via backpropagation, i.e, 
$
     \vtheta_{k+1} = \vtheta_k - \eta \vd_k = \vtheta - \eta \frac{\partial g_\vlam(\vL(\vtheta))}{\partial \vtheta} |_{\vtheta_k},   
$
where $\vd_k$ is called the updating direction at the $k$th iteration.

\paragraph{Gradient manipulation-based methods.}
Besides, directly optimizing the aggregation function, a number of so-called \emph{``gradient manipulation methods''} solve an updating direction $\vd_k$ using gradient information for each iteration for some specific purpose. For example, as listed in~\Cref{tab:finite_solver}, EPO~\cite{mahapatra2020multi} aims to find ``exact Pareto solutions'' (intersection points of Pareto front and preference vectors), HVGrad~\cite{deist2021multi} aim to maximize the hypervolume of a set of solutions, MOO-SVGD~\cite{liu2021profiling} aim to find diverse solutions, and PMTL~\cite{lin2019pareto} aim to find sector-constrained Pareto solutions.
\begin{wraptable}{r}{0.6\textwidth}
\setlength\tabcolsep{1.5 pt}
\centering
\small
\vspace{-10pt}
\begin{threeparttable}
\caption{\footnotesize MOO solvers, properties, and complexities.}
\label{tab:finite_solver}
    \tiny
    \begin{tabular}{llll}
    \toprule
    \textbf{Method} & \textbf{Solution Property} & \textbf{Complexity} & \textbf{Pref.} \\
    \midrule
    \href{https://proceedings.mlr.press/v119/mahapatra20a.html}{EPO}~\cite{mahapatra2020multi} & Exact solutions &  $O(m^2nK)$ & \checkmark \\
    \href{https://arxiv.org/abs/2102.04523}{HVGrad}~\cite{deist2021multi} & Solutions with maximal HV & $O(m^2nK^2)$ & \texttimes \\
    \href{https://arxiv.org/abs/1810.04650}{MGDA-UB}~\cite{sener2018multi} & Random solutions & $O(m^2nK)$ & \texttimes  \\
    \href{https://papers.nips.cc/paper_files/paper/2021/hash/7bb16972da003e87724f048d76b7e0e1-Abstract.html}{MOO-SVGD}~\cite{liu2021profiling} & Diversity by particles repulsion & $O(m^2nK^2)$ & \texttimes \\
    
    \href{https://arxiv.org/abs/2402.09492}{PMGDA}~\cite{zhang2024pmgda} & Solutions under specific demands & $O(m^2nK)$ & \checkmark \\

    \href{https://arxiv.org/abs/1912.12854}{PMTL}~\cite{lin2019pareto} & Solutions in sectors & $O(m^2nK^2)$ & \texttimes \\
    
    \href{https://arxiv.org/abs/2111.10603}{Random}~\cite{lin2021reasonable} & Random solutions & $O(m^2nK)$ & \texttimes \\
        \href{https://link.springer.com/book/10.1007/978-1-4615-5563-6}{Agg-LS}~\cite{miettinen1999nonlinear} & Convex part of a PF & $O(mnK)$ & \checkmark \\
    
    \href{https://ieeexplore.ieee.org/document/4358754}{Agg-Tche}~\cite{zhang2007moea} & Exact solutions &  $O(mnK)$ & \checkmark \\
    
    \href{https://ieeexplore.ieee.org/document/7927726}{Agg-mTche}~\cite{ma2017tchebycheff} & Exact solutions & $O(mnK)$ & \checkmark \\
    
    \href{https://ieeexplore.ieee.org/document/4358754}{Agg-PBI}~\cite{zhang2007moea} & Approximate exact solutions & $O(mnK)$ & \checkmark \\
    
    \href{https://arxiv.org/abs/2103.13392}{Agg-COSMOS}~\cite{ruchte2021scalable}  & Approximate exact solutions & $O(mnK)$ & \checkmark \\
    
    \href{https://arxiv.org/abs/2402.19078}{Agg-SmoothTche}~\cite{lin2024smooth} & Approximate exact solutions & $O(mnK)$ & \checkmark \\
    \bottomrule
    \end{tabular}
    \begin{tablenotes}
    \tiny 
       \item \tiny $m$: number of objectives. $n$: number of decision variables. $K$: number of subproblems. $m$ is usually small (e.g., 2-4), $K$ is relatively large (e.g., 20-40), and $n$ is particularly large (e.g., 10,000). Therefore, $m^2$ is not a big concern, while $K^2$ and $n^2$ are big concerns. Complexity is for time complexity, and Pref. denotes whether this method is preference-based or not. 
    \end{tablenotes}
\end{threeparttable}
\vspace{-10pt}
\end{wraptable}
Interestingly, until now, all these gradient manipulation methods can be implemented in two steps: \textcircled{1} calculating a dynamic weight vector \( \tilde{\bm \alpha}\) and then \textcircled{2} performing backpropagation on a generalized aggregation function \( \tilde{g}_\vlam(\vL(\vtheta_k)) \), where \( \tilde{g}_\vlam(\vL(\vtheta_k)) = \sum_{i=1}^m \tilde{\alpha}_i L_i(\vtheta)\). At each iteration, gradient manipulation methods can be equivalently expressed as updating the gradient of its induced generalization aggregation function $\tilde{g}_\vlam(\vL(\vtheta_k))$. The weight vector $\tilde{\bm \alpha}$ are achieved by solving a linear programming (LP) problem (e.g., \cite{mahapatra2020multi}[Eq. 24]) in EPO, a quadratic programming (QP) problem (e.g., \cite{lin2019pareto}[Eq. 14]) in PMTL, or other more complex algorithms as used to calculate the hypervolume gradient~\cite{emmerich2014time}.

Some MOO solvers accept preference vectors $\vlam$ as input, termed \emph{preference-based}, affecting Pareto solution positions. The others, called \emph{preference-free}, do not accept preferences, such as those maximizing dominated hypervolume. A summary of these solvers is in~\Cref{tab:finite_solver}.

\subparagraph{Zero-order optimization.}
For the previous discussions, we assume that all gradients of objective functions $\nabla L_i(\vtheta)$ can be easily achieved via backward propagation. However, for some black-box optimization problems, $\nabla L_i(\vtheta)$'s may not easily be achieved. Therefore, \algoname~not only supports first-order optimization, but also supports zero-order optimization methods with estimated gradients $\hat{\nabla} L_i(\vtheta)$ such as evolutionary strategy (ES)~\cite{beyer2002evolution}.

\subsection{Pareto set learning solvers} \label{sec:psl}

\begin{table}[h!]

\caption{Comparison of different PSL methods.} 
\label{tab:psl_comparison}
\centering
\small
\begin{threeparttable}
    \begin{tabular}{llll}
    \toprule
    \textbf{Method} & \textbf{Property} & Vector $\tilde{\bm \alpha}$ & Matrix $\mB$ \\
    \midrule
    \href{https://openreview.net/pdf/9c01e8c47f7e80e87af0175ac2a5e9a356f518bd.pdf}{EPO-based PSL}~\cite{navon2020learning} & Exact solutions & MOO solvers & BP\\
    
    \href{https://arxiv.org/abs/2402.09492}{PMGDA-based PSL}~\cite{zhang2024pmgda} & Solutions under specific demands & MOO solvers & BP \\\href{https://openreview.net/pdf/9c01e8c47f7e80e87af0175ac2a5e9a356f518bd.pdf}{Aggregation-based PSL}~\cite{navon2020learning} & Solutions with optimal aggregation values & BP & BP\\
    
    \href{https://arxiv.org/pdf/2310.20426}{Evolutionary PSL}~\cite{lin2023evolutionary} & Mitigating local minima by ES & BP & ES \\
    
    \href{https://openreview.net/pdf?id=a2uFstsHPb}{LoRA PSL}~\cite{chen2024efficient,zhong2024panacea,dimitriadis2024pareto} & A lighter Pareto model structure & BP & BP \\
    \bottomrule
    \end{tabular}
    \begin{tablenotes}
        \footnotesize
        \item 
        BP: backward propagation, 
        ES: evolutionary strategy.
    \end{tablenotes}
\end{threeparttable}
\end{table}

\algoname~supports Pareto Set Learning (PSL), which trains a model with parameter $\vphi$ to approximate the \emph{entire} PS/PF. A Pareto model is denoted as $\vtheta_\vphi(\cdot) : \mDelta_m \mapsto \mathbb{R}^n$ with input as a preference vector and output as a Pareto solution. 

\begin{wrapfigure}{r}{0.6\textwidth}
    \centering
    \vspace{-15pt}
    \subfloat[Synthetic problem]{\includegraphics[width=0.3\textwidth]{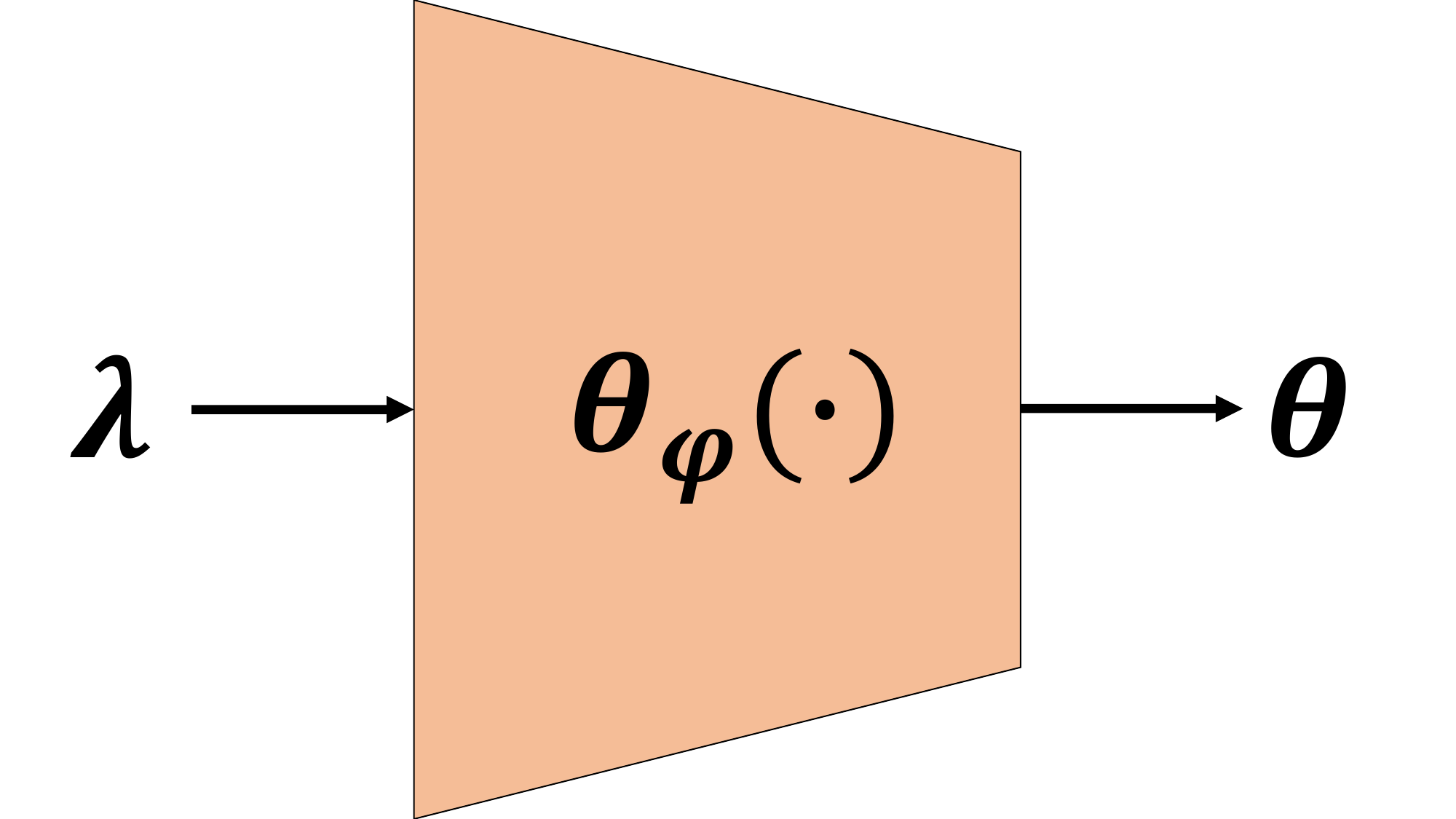}\label{<figure1>}}
    \subfloat[MultiTask Learning]{\includegraphics[width=0.3\textwidth]{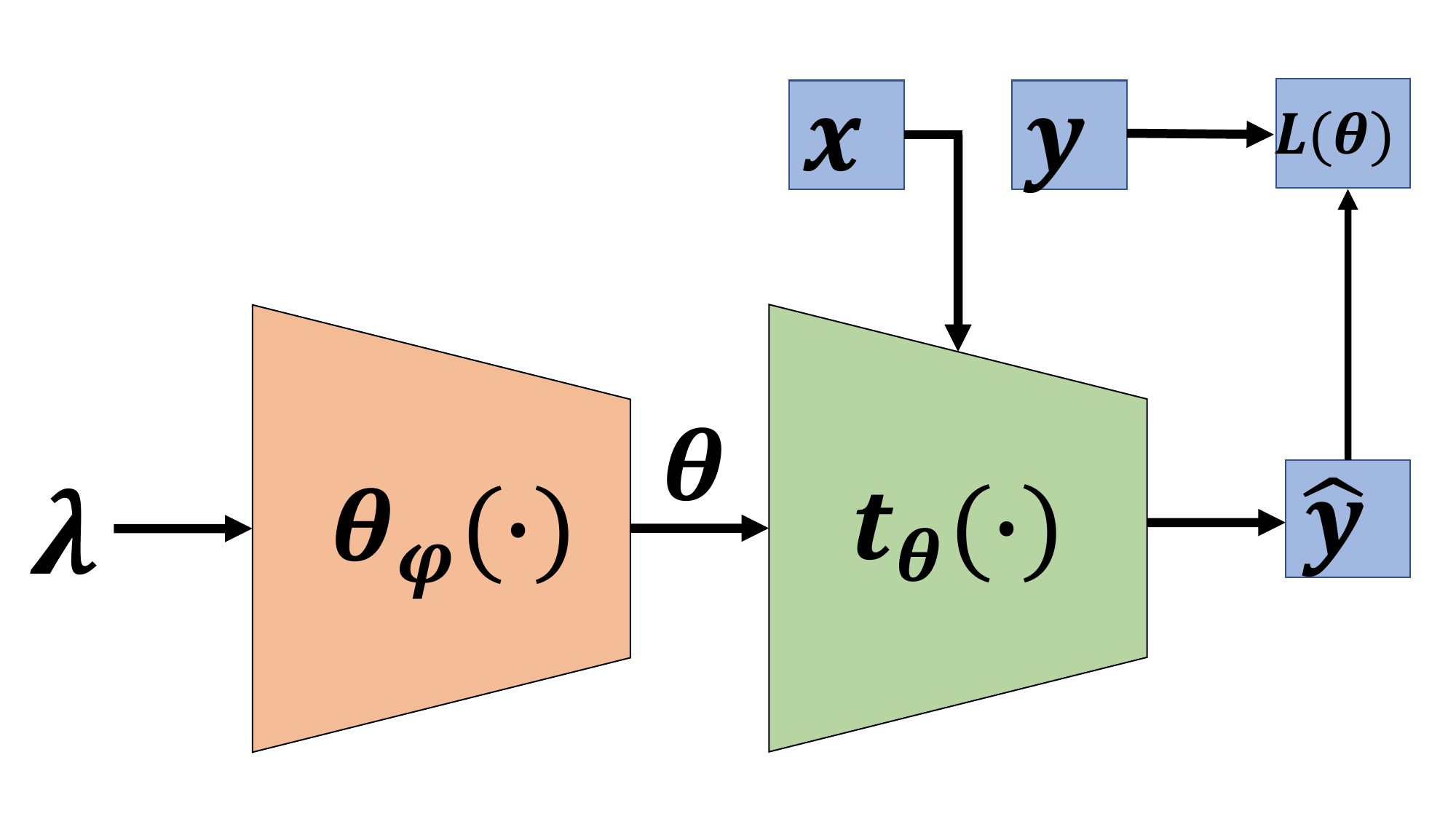}\label{<figure2>}}
    \caption{Architecture of Pareto models.}
    \vspace{-15pt}
    \label{fig:psl:arch}
\end{wrapfigure}

\textbf{PSL Architecture.} Pareto models vary in structure. For synthetic problems, the simplest model is a fully-connected neural network that takes a preference as input and outputs the corresponding Pareto solution. In multitask learning, the input is a pair $(\vx, \vy)$ from dataset $\mathcal{D}$, and the decision variable $\vtheta$ represents the target network's parameter, with $\phi$ as the hypernetwork's parameter~\cite{ha2016hypernetworks}. The loss vector is calculated as $\vL(\vtheta) = \E_{(\vx,\vy)\sim \mathcal{D}}(\ell(\bm t_\vtheta(\vx), \vy))$, where $\ell$ is a basic loss function like cross-entropy or mean square error. Structures of these two models are illustrated by \Cref{fig:psl:arch}. Besides these two models, \algoname~also supports LoRA (low rank adaptation)-based PSL~\cite{chen2024efficient,zhong2024panacea,dimitriadis2024pareto}, which admits a low rank adaptation structure and other structures. PSL structures are decoupled from the training loss and used as plug-ins.

\textbf{PSL Training.} Goal of PSL is to find a model with parameter $\vphi$ optimizing the PSL loss $\ell_\mathrm{psl}$
$
    \min_\vphi \ell_\mathrm{psl} = \E_{\vlam \sim \mathrm{Dir}(\bm p)} \tilde{g}_\vlam(\vL( \vtheta_\vphi(\vlam) )),
$
where $\mathrm{Dir}(\bm p)$ is a Dirichlet distribution with hyperparameter $\bm p$. $\tilde{g}_\vlam(\cdot)$ can either be a generalized aggregation function as introduced in the previous section or a normal aggregation function. The gradient of $\ell_\mathrm{psl}$ can be estimated by the chain rule:

\begin{equation}
    \underbrace{\frac{\partial \ell_\mathrm{psl}}{\partial \vphi}}_{1 \times D}
    = 
    \E_{\vlam \sim \mathrm{Dir}(\bm p)} \underbrace{\grad{\tilde{g}_\vlam}{\vL}}_{\tilde{\bm \alpha}: (1 \times m)} \cdot \underbrace{\grad{\vL}{\vtheta}}_{\mB: (m \times n)} \cdot \underbrace{\grad{\vtheta}{\vphi}}_{\mC: (n \times D)}.
    \label{eqn:grad:psl}
\end{equation}

Empirically, the expectation involved in \Cref{eqn:grad:psl} is estimated using a mini-batch of \(K\) preferences. The gradient vector \(\tilde{\bm \alpha} =(\tilde{\alpha}_1, \ldots, \tilde{\alpha}_m)\) is computed by MOO solvers as introduced in the previous section. The gradient matrix \(\mB\) can be calculated either via backpropagation (when gradients are easily obtained) or using a zero-order method such as ES. \(\mC\) can always be estimated through backward propagation, since \(\vtheta\) is a continuous vector function of \(\vphi\). Gradient calculations in existing PSL methods is summarized in \Cref{tab:psl_comparison}. Parameter \(\vphi\) is iteratively updated by gradient descent: \(\vphi \xleftarrow{} \vphi - \eta \frac{\partial \ell_\mathrm{psl}}{\partial \vphi}\).

\subsection{Multiobjective Bayesian optimization solvers}
\label{sec:mobod}
When the evaluation of objective functions is costly, multiobjective Bayesian optimization (MOBO) is often the preferred methodology for tackling such challenges. While there are several existing libraries, such as BoTorch~\cite{balandat2020botorch} and HEBO~\cite{Cowen-Rivers2022-HEBO}, that facilities MOBO, they largely overlook algorithms that leverage decomposition techniques like PSL. To bridge this gap, \algoname~also includes three decomposition-based MOBO algorithms, including PSL-MOBO~\cite{lin2022pslmobo}, PSL-DirHVEI~\cite{lin2022pslmobo,zhao2024hypervolume}, and DirHV-EGO~\cite{zhao2024hypervolume}. In each iteration, these methods build Gaussian process (GP) models for each objectives and generate a batch of query points for true function evaluations.

PSL-MOBO is an extension of PSL method for expensive MOPs. It optimizes the preference-conditional acquisition function (AF) $\alpha(\vtheta|\vlam)$ over an infinite number of preference vectors to generate a set of promising candidates:
$
   \min_\vphi \ell_\mathrm{psl} = \E_{\vlam \sim \mathrm{Dir}(\bm p)} \mbr{\alpha(\vtheta_\vphi(\vlam)|\vlam)}, \vtheta_\vphi(\vlam) : \mDelta_m \mapsto\R^N.
$
 \begin{wraptable}{r}{0.6\textwidth}
     \centering
     \setlength\tabcolsep{1.5 pt}
     \caption{\textbf{Supported preference-conditional AFs}.} \label{tab:mobod:acquisition}
     \small
     \begin{tabular}{c|c}
     \toprule
      & preference-conditional AFs  \\ \midrule
      TLCB & $\alpha_\mathrm{{TLCB}}(\vtheta|\vlam)=\underset{i \in [m]}{\max}\{\lambda_i(\hat{\mu}_i(\vtheta)-\beta\hat{\sigma}_i(\vtheta)-z_i^*)\}$  \\
         DirHV-EI &  $\alpha_\mathrm{{DirHVEI}}(\vtheta | \vlam) =\mathbb{E}_{\vy\sim p(\vy | \vtheta,\mathcal{D})} \mbr{ \prod_{i=1}^{m} [\xi_i-y_i]_{+} }$ \\
     \bottomrule
\end{tabular}
\vspace{-9pt}
\end{wraptable}

Currently, our library supports two representative preference-conditional AFs: the Tchebycheff scalarization of lower confidence bound (TLCB)~\cite{lin2017cec,lin2022pslmobo,paria2020flexible} and the expected direction-based hypervolume improvement (DirHV-EI)~\cite{zhao2024hypervolume}. We note that DirHV-EI can be regarded as an unbiased estimation of a weighted expected hypervolume improvement. Our library also supports DirHV-EGO, which employs a finite set of predetermined reference vectors as in~\cite{zhang2007moea}.


\section{Empirical studies} \label{sec:results}
In this section, we present the empirical results of \algoname. The experiments were conducted on a personal computer with an i7-10700 CPU and a 10GB RTX3080 GPU. GPU acceleration analysis was performed using RTX3080, 4060, and 4090. The results cover six areas: MOO solvers for synthetic problems, Pareto set learning solvers for synthetic problems, MOO solvers for MTL problems, Pareto set learning solvers for MTL problems, MOBO for synthetic problems, and GPU acceleration (details in \Cref{sec:gpu}).

\subsection{MOO solvers for synthetic problems}
We report the performance of various finite Pareto solution solvers using the VLMOP2 problem, where $f_1(\vtheta) = 1 - \exp (-\sum_{i=1}^{n} (\theta_i - \frac{1}{\sqrt{n}} )^2 )$, $f_2(\vtheta) = 1 - \exp (-\sum_{i=1}^{n} (\theta_i + \frac{1}{\sqrt{n}})^2)$. VLMOP2 has been widely studied in the literature~\cite{mahapatra2020multi,lin2019pareto} since its PF is non-convex~\footnote{In MOO, a PF is convex or non-convex based on whether the objective space is convex or non-convex. The PF represents the non-dominated boundary of the objective space.}. Some methods (e.g., Agg-LS) fail immediately on this problem since its PF is non-convex. Visualization results are shown by \Cref{fig:finit_synthetic} and the numerical results are shown by \Cref{tab:finite_synthetic}. We present the following key findings:

\begin{figure} %
    \centering
    \subfloat[Agg-COSMOS]{\includegraphics[width= \pentawidth \textwidth]{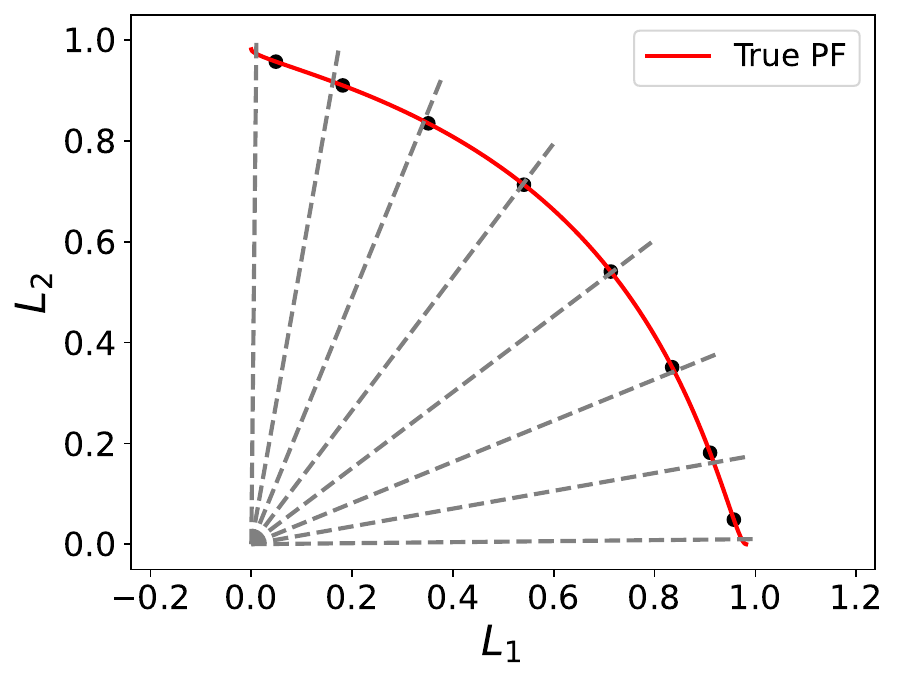}}
    \subfloat[Agg-LS]{\includegraphics[width= \pentawidth \textwidth]{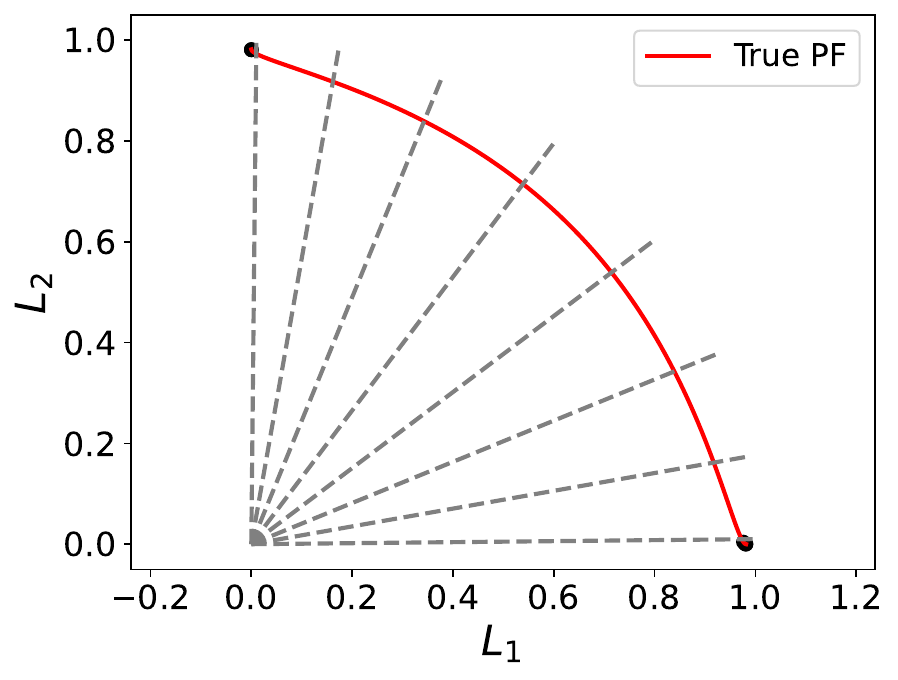}} 
    \subfloat[Agg-PBI]{\includegraphics[width= \pentawidth \textwidth]{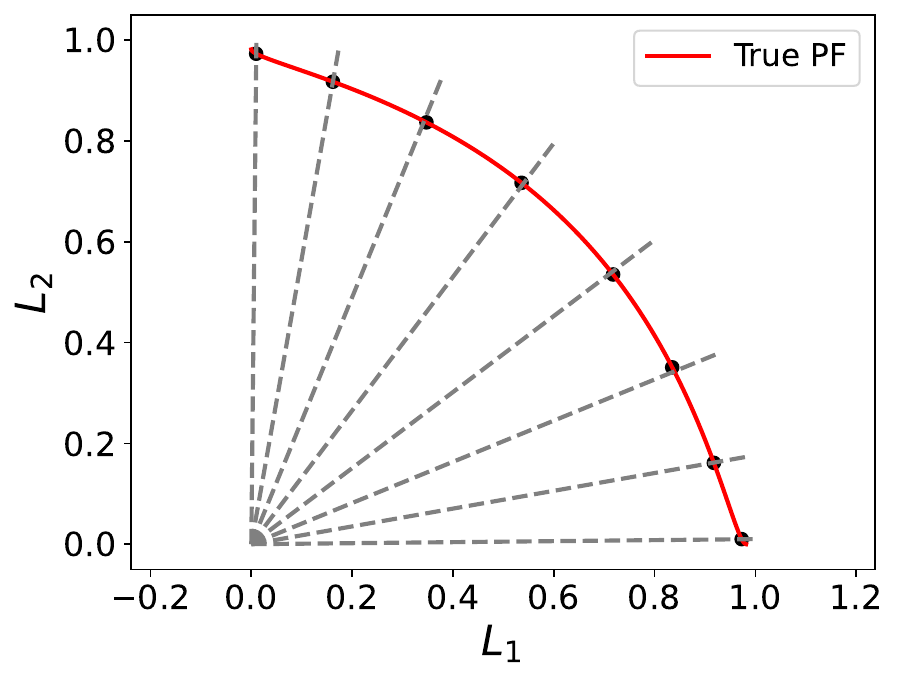}}
    \subfloat[Agg-SmoothTche]{\includegraphics[width= \pentawidth \textwidth]{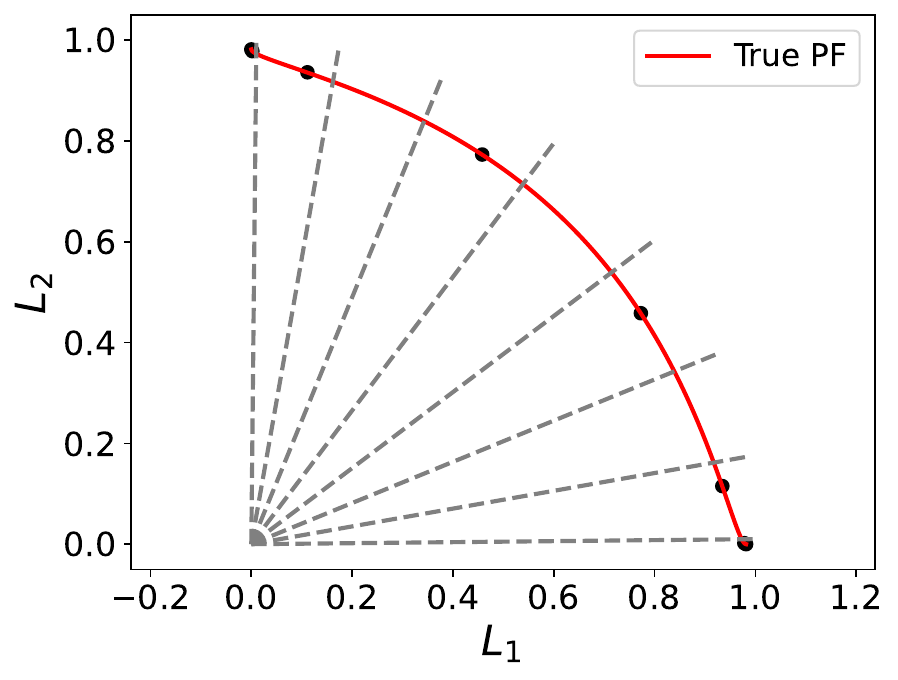}}
    \subfloat[Agg-Tche]{\includegraphics[width= \pentawidth \textwidth]{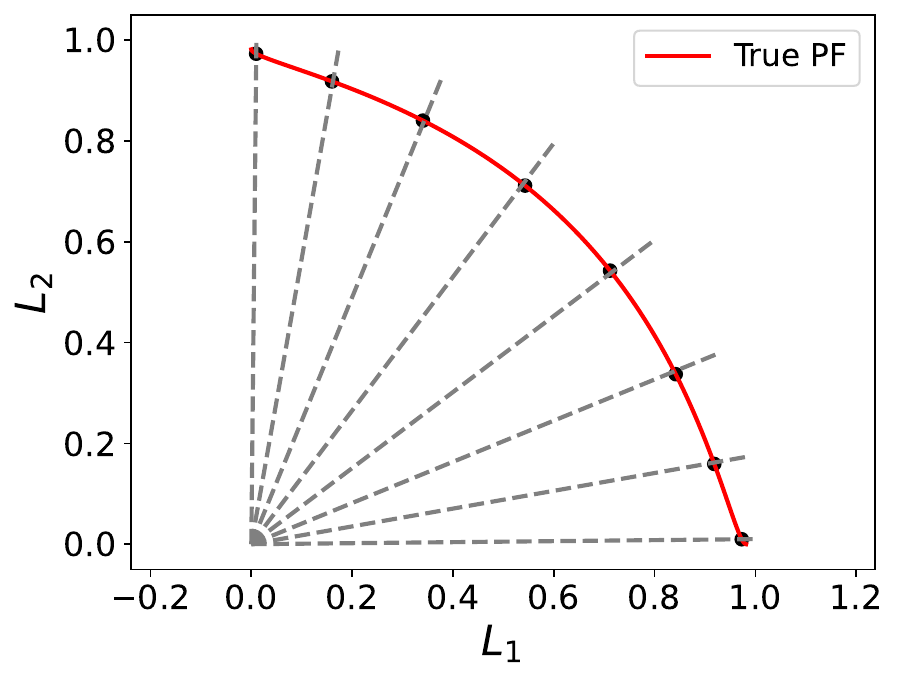}} \\
    \vspace{-10pt}
    \subfloat[EPO]{\includegraphics[width= \pentawidth \textwidth]{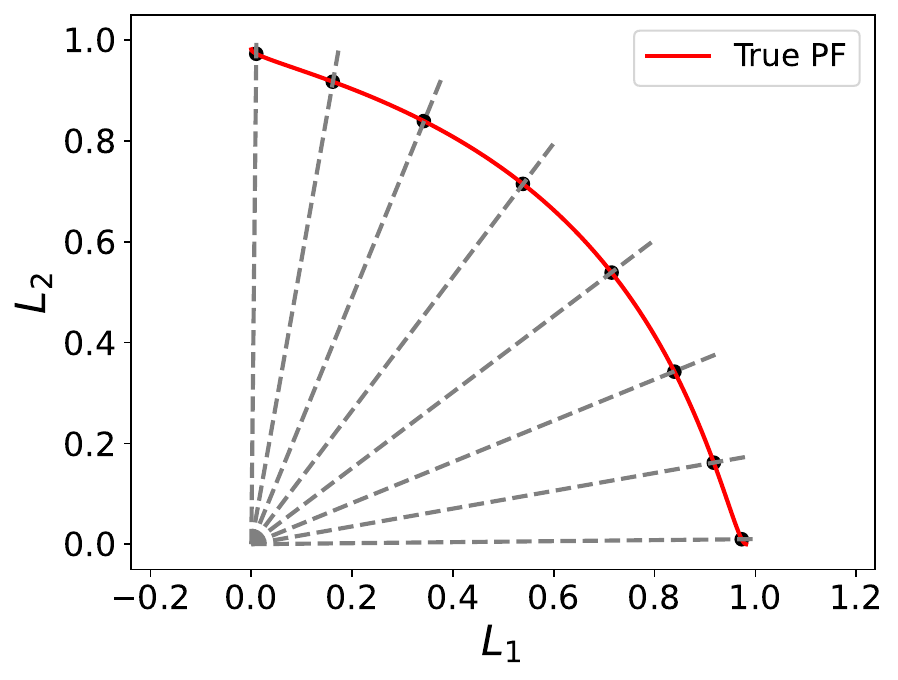}} 
    \subfloat[HVGrad]{\includegraphics[width= \pentawidth \textwidth]{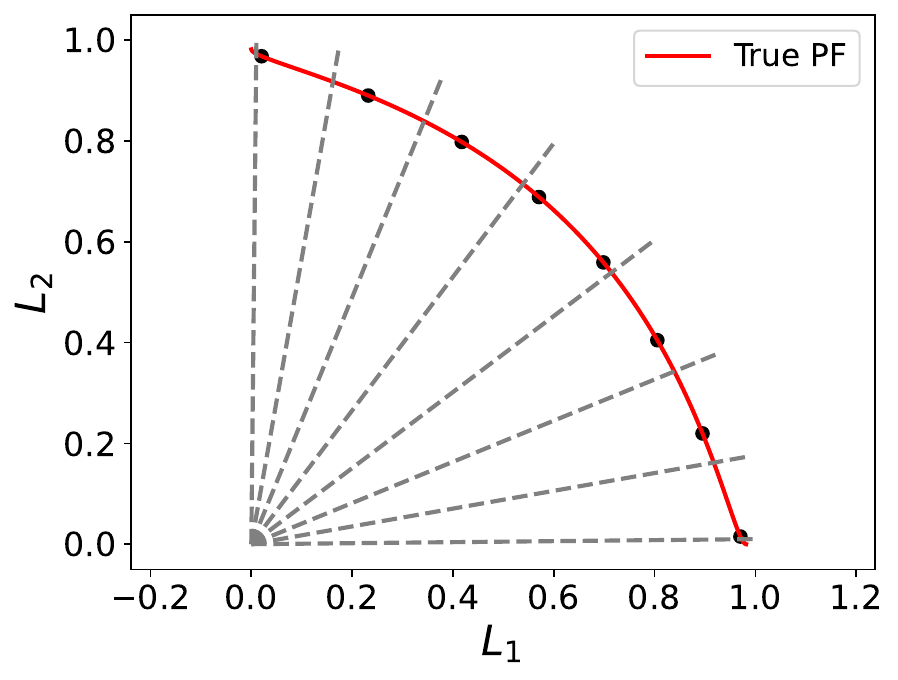}}
    \subfloat[MGDA-UB]{\includegraphics[width= \pentawidth \textwidth]{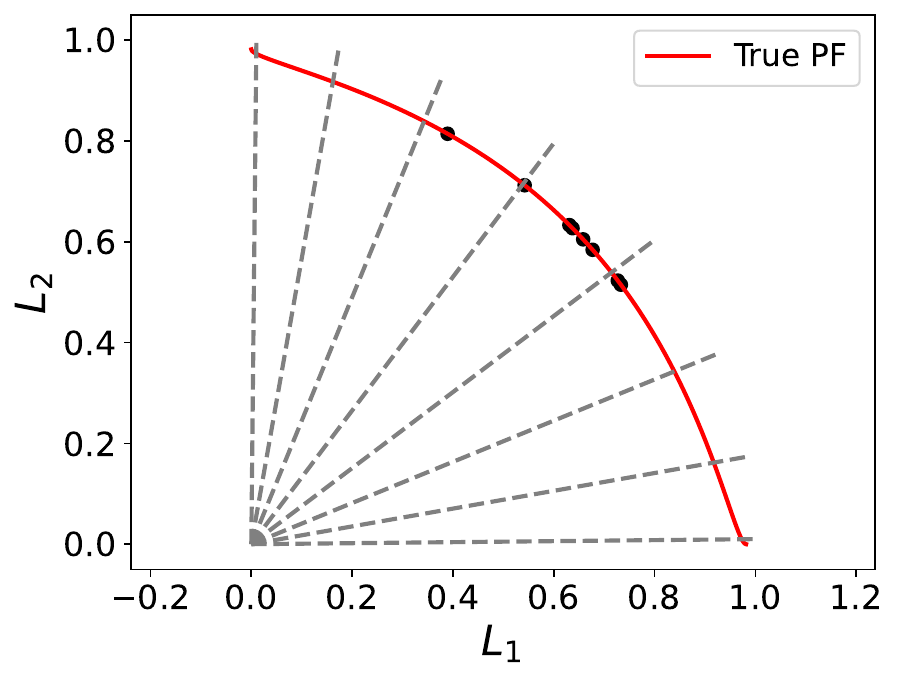}}
    \subfloat[MOO-SVGD]{\includegraphics[width= \pentawidth \textwidth]{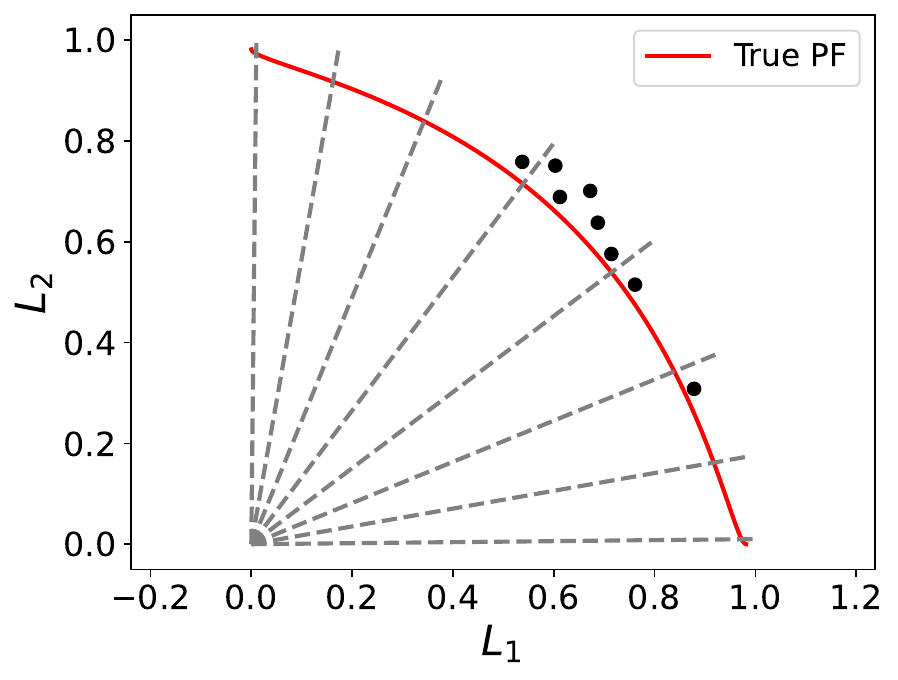}}
    \subfloat[PMTL]{\includegraphics[width= \pentawidth \textwidth]{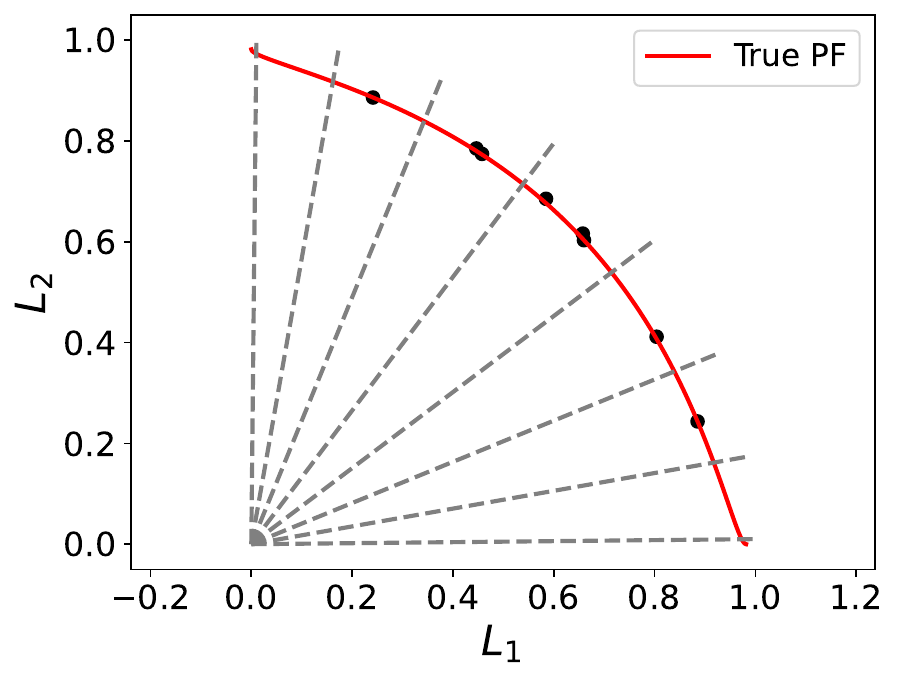}}
    \\
    \caption{Finite Pareto solutions by ten MOO solvers on VLMOP2 problem.} \label{fig:finit_synthetic}
\end{figure}

\begin{table}[ht]
\setlength\tabcolsep{2.0 pt}
\centering
\tiny
\begin{threeparttable}
\caption{Numerical results of finite Pareto solutions for the VLMOP2 problem.}
\label{tab:finite_synthetic}
    \begin{tabular}{l|l|l|l|l|l|l|l|l}
    \toprule
    Method & $l_\mathrm{min}$ & $sl_\mathrm{min}$ & Spacing & Sparsity & HV & IP & Cross Angle & PBI \\
    \midrule
    EPO & 0.162 (0.000) & 0.061 (0.000) & 0.029 (0.001) & 0.043 (0.000) & 0.283 (0.000) & 0.776 (0.000) & \textbf{0.046 (0.041)} & \textbf{0.930 (0.003)} \\
    MGDA-UB & 0.012 (0.013) & -0.098 (0.011) & 0.036 (0.011) & \textbf{0.006 (0.001)} & 0.228 (0.008) & 0.606 (0.010) & 31.278 (1.533) & 2.986 (0.088) \\
    PMGDA & 0.150 (0.001) & 0.055 (0.000) & 0.034 (0.001) & 0.042 (0.000) & 0.283 (0.000) & 0.775 (0.000) & 0.318 (0.037) & 0.952 (0.003) \\
    Random & 0.000 (0.000) & -0.161 (0.002) & \textbf{0.000 (0.000)} & 0.272 (0.000) & 0.044 (0.000) & 0.410 (0.127) & 52.290 (12.938) & 3.894 (0.590) \\
    MOO-SVGD & 0.060 (0.002) & -0.077 (0.004) & 0.033 (0.018) & 0.009 (0.003) & 0.212 (0.003) & 0.633 (0.024) & 29.647 (3.305) & 2.963 (0.197) \\
    PMTL & 0.014 (0.010) & -0.068 (0.009) & 0.061 (0.020) & 0.018 (0.007) & 0.260 (0.012) & 0.706 (0.004) & 15.036 (1.270) & 1.993 (0.093) \\
    HVGrad & \textbf{0.182 (0.000)} & \textbf{0.067 (0.000)} & 0.016 (0.000) & 0.041 (0.000) & \textbf{0.286 (0.000)} & 0.578 (0.069) & 34.090 (8.607) & 3.062 (0.465) \\
    Agg-LS & 0.000 (0.000) & -0.159 (0.001) & 0.002 (0.001) & 0.272 (0.001) & 0.043 (0.002) & \textbf{0.227 (0.008)} & 71.168 (0.958) & 4.764 (0.047) \\
    Agg-Tche & 0.158 (0.001) & 0.061 (0.000) & 0.031 (0.001) & 0.043 (0.000) & 0.283 (0.000) & 0.348 (0.000) & 55.174 (0.049) & 3.889 (0.004) \\
    Agg-PBI & 0.113 (0.074) & 0.032 (0.046) & 0.045 (0.030) & 0.042 (0.002) & 0.281 (0.002) & 0.657 (0.097) & 11.374 (9.125) & 1.434 (0.402) \\
    Agg-COSMOS & 0.141 (0.000) & 0.045 (0.000) & 0.035 (0.000) & 0.039 (0.000) & 0.285 (0.000) & 0.771 (0.000) & 1.085 (0.000) & 1.011 (0.000) \\
    Agg-SmoothTche & 0.004 (0.000) & -0.074 (0.000) & 0.154 (0.001) & 0.074 (0.000) & 0.244 (0.000) & 0.276 (0.000) & 63.106 (0.018) & 4.253 (0.001) \\
    \bottomrule
    \end{tabular}
    \begin{tablenotes}
    \item All methods were run five times with random seeds; results are presented in the format of (mean)(std).
    \end{tablenotes}
\end{threeparttable}
\end{table}

\textcircled{1}. Agg-COSMOS produces rough `exact' Pareto solutions due to a cosine similarity term encouraging Pareto objectives to be close to preference vectors. However, the position of Pareto objectives can not be determined. Agg-LS can only find two endpoints of the PF. 
\textcircled{2}. Agg-PBI generates ``exact'' Pareto solutions when the coefficient of \( d_2 \) exceeds a specific value~\cite{wang2021parameter}. However, this parameter is challenging to tune, which is influenced by the curvature of a PF. Additionally, PBI may transform a convex multi-objective optimization problem into a non-convex one, making Agg-PBI less recommended.
\textcircled{3}. Agg-Tche generates diverse solutions and produces exact Pareto solutions corresponding to the element-wise inverse of the preference vector. Both Agg-Tche and Agg-SmoothTche retain convexity - their aggregation functions keep convex when all objectives are convex.
\textcircled{4}. HVGrad updates the decision variable using the hypervolume gradient, resulting in the largest hypervolume.
PMTL is a two-stage method. In the first stage, solutions are updated to specific regions and in the second stage, solutions are updated to Pareto solutions constrained in these specific regions. MOO-SVGD's update direction has two conflicting goals: promoting diversity and ensuring convergence. This conflict makes MOO-SVGD less stable and take more iterations to converge. 
\textcircled{5}. Among these methods, MGDA-UB, Random, Agg-PBI, and MOO-SVGD exhibit relatively large deviations. MGDA-UB and Random generate arbitrary Pareto solutions due to their computation nature. Agg-PBI results in non-convex aggregation functions, leading to variable solutions based on different initializations.

\textbf{Conclusion.} For synthetic problems, the most recommended method is \emph{Agg-Tche} since (1) it keeps convexity, (2) it finds `exact' Pareto solutions under quite mild conditions (3) it does not need to calculate the Jacobian matrix for each iteration.

\subsection{Pareto set learning on synthetic problems}
\begin{figure}%
    \centering
    \subfloat[COSMOS]{\includegraphics[width= \sixwidth \textwidth]{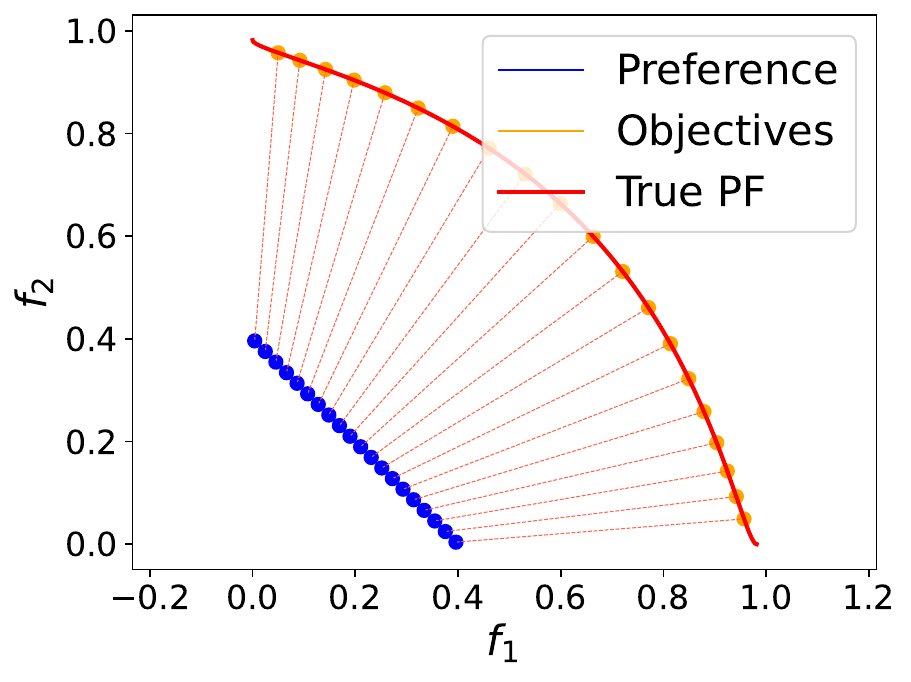}}
    \subfloat[Agg-LS]{\includegraphics[width= \sixwidth \textwidth]{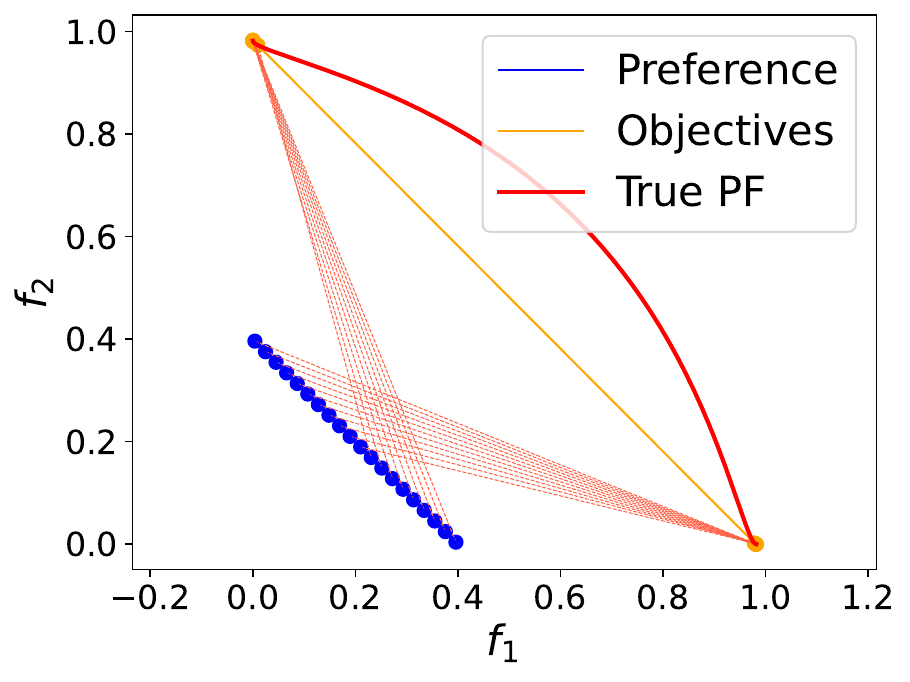}} 
    \subfloat[SmoothTche]{\includegraphics[width= \sixwidth \textwidth]{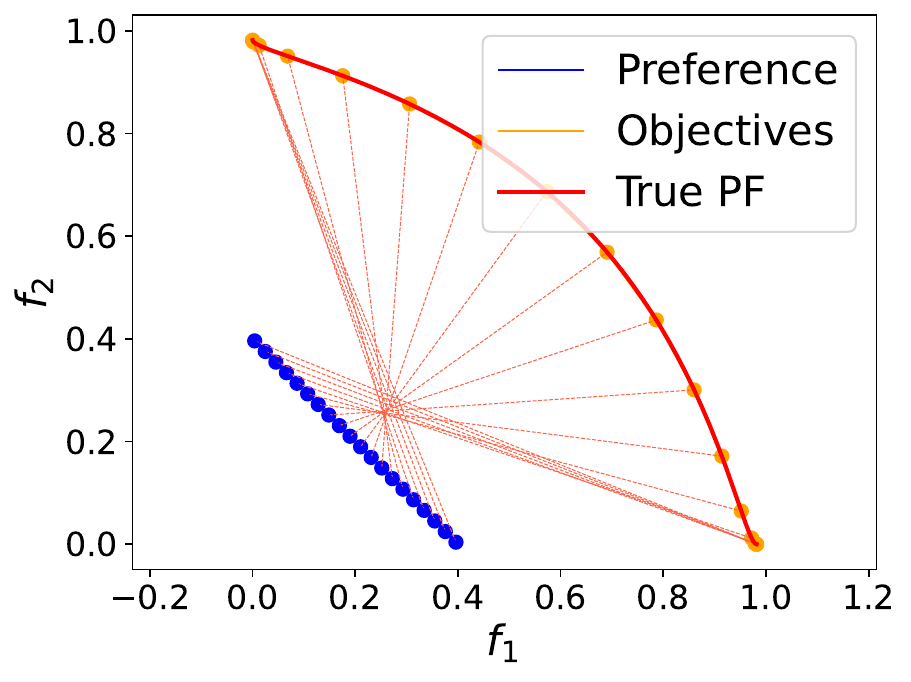}}
    \subfloat[Tche]{\includegraphics[width= \sixwidth \textwidth]{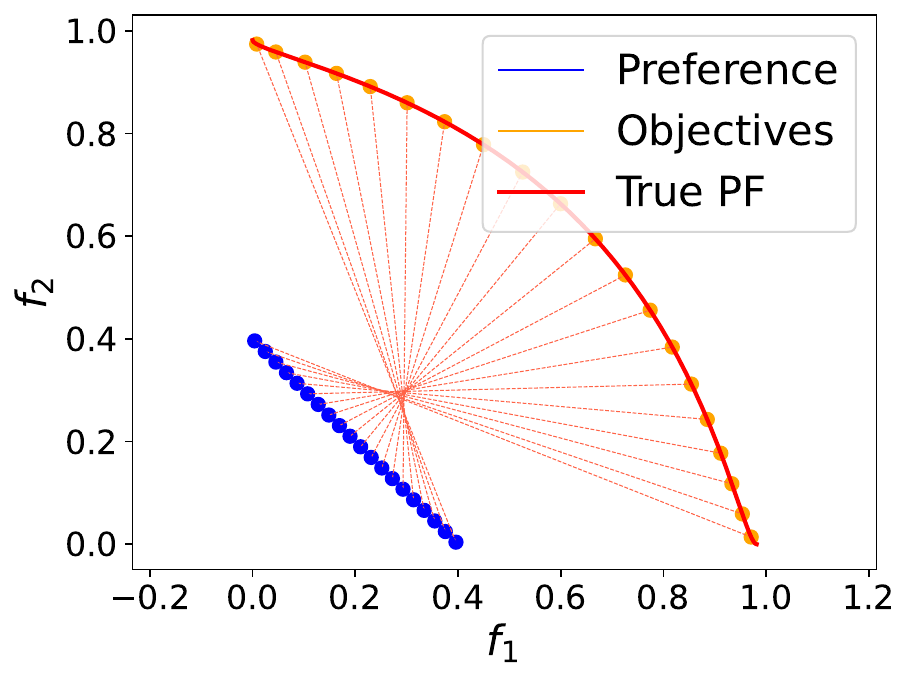}}
    \subfloat[EPO]{\includegraphics[width= \sixwidth \textwidth]{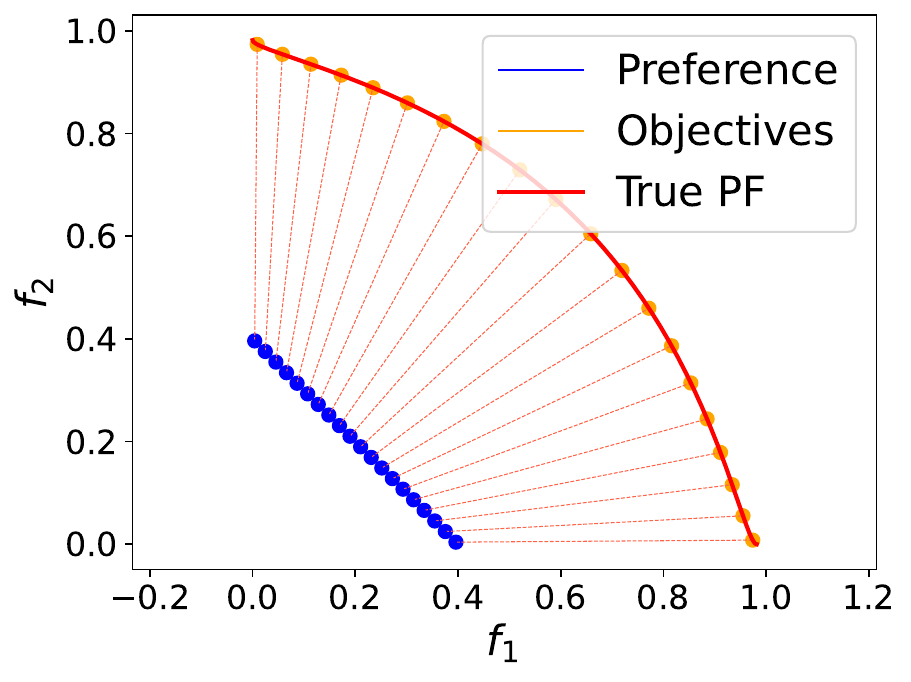}} 
    \subfloat[PMGDA]{\includegraphics[width= \sixwidth \textwidth]{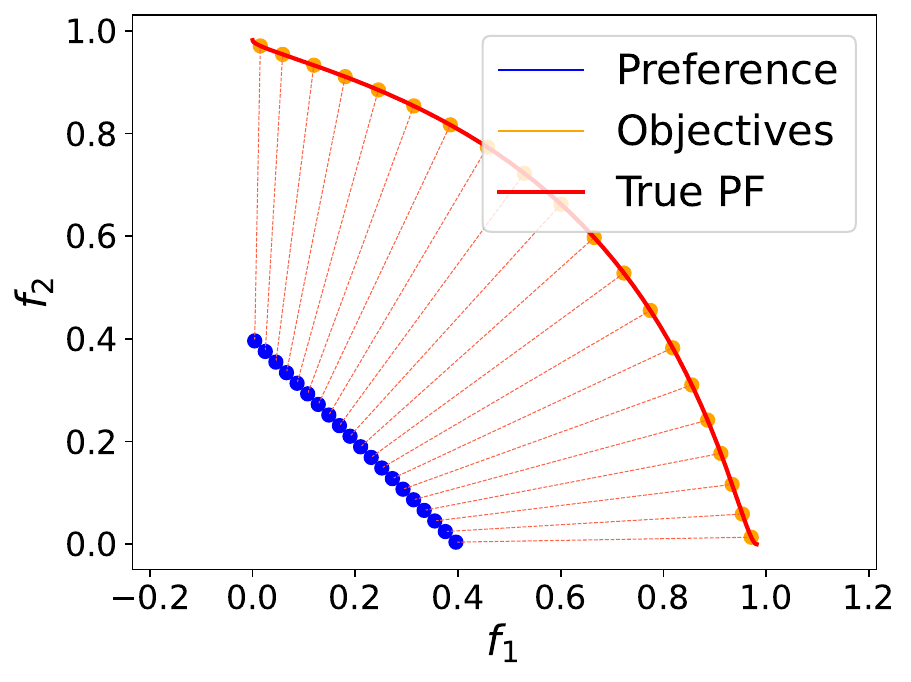}}
    \\
    \caption{Predicted Pareto solutions by different PSL solvers on VLMOP2 problem.} \label{fig:psl_synthetic}
\end{figure}

In this section, we present PSL results (\Cref{fig:psl_synthetic,tab:psl_synthetic}) also using VLMOP2. We have:

\begin{table}[]
\setlength\tabcolsep{2.0 pt}
\centering
\scriptsize
\caption{Pareto set learning results on VLMOP2 problem.} \label{tab:psl_synthetic}
\tiny 
    \begin{tabular}{l|l|l|l|l|l|l|l|l|l}
    \toprule
    Method & $l_\mathrm{min}$ & $sl_\mathrm{min}$ & Spacing           & Sparsity          & HV                & IP                & Cross Angle       & PBI               & Span              \\
    \midrule
    COSMOS & 0.045 (0.000)     & -0.127 (0.000)     & 1.560 (0.004)     & \textbf{0.525 (0.000)} & 0.318 (0.000)     & 0.752 (0.000)     & 0.950 (0.001)     & 0.995 (0.000)     & 0.907 (0.000)     \\
    Agg-LS & 0.000 (0.000)     & -0.258 (0.000)     & \textbf{0.115 (0.094)} & 13.245 (1.871)    & 0.045 (0.003)     & \textbf{0.239 (0.001)} & 70.541 (0.156)    & 4.811 (0.007)     & \textbf{0.982 (0.000)} \\
    Agg-Tche & 0.047 (0.004)     & -0.121 (0.000)     & 1.476 (0.146)     & 0.579 (0.008)     & 0.319 (0.000)     & 0.383 (0.003)     & 51.558 (0.280)    & 3.775 (0.011)     & 0.955 (0.005)     \\
    SmoothTche & 0.000 (0.000)     & -0.187 (0.000)     & 6.711 (0.003)     & 1.060 (0.000)     & 0.302 (0.000)     & 0.300 (0.000)     & 60.386 (0.001)    & 4.169 (0.000)     & 0.982 (0.000)     \\
    EPO & \textbf{0.050 (0.001)} & \textbf{-0.120 (0.000)} & 1.332 (0.078)     & 0.583 (0.005)     & 0.319 (0.000)     & 0.756 (0.000)     & 0.388 (0.098)     & 0.952 (0.008)     & 0.961 (0.003)     \\
    PMGDA & 0.047 (0.000)     & -0.121 (0.000)     & 1.446 (0.051)     & 0.580 (0.002)     & \textbf{0.319 (0.000)} & 0.756 (0.000)     & \textbf{0.215 (0.062)} & \textbf{0.939 (0.005)} & 0.958 (0.001) \\
    \bottomrule
    \end{tabular}
\end{table}

\textcircled{1}. PSL with the COSMOS aggregation function fails to find all marginal Pareto solutions because COSMOS does not guarantee the discovery of the entire PS/PF. PSL with linear scalarization function could not fit the two endpoints of the PF. Those PSL results inherit from their base MOO solvers.
\textcircled{2}. PSL with the smooth Tchebycheff function finds diverse but non-exact Pareto solutions. In contrast, PSL with Agg-Tche, EPO, and PMGDA as base solvers discovers the entire PS/PF, as all three methods find exact Pareto solutions. By traversing the preference simplex, the model accurately fits the entire PS.

\textbf{Conclusion.} The most recommended method is still \emph{Agg-Tche-based PSL} since its basic MOO solver Agg-Tche has attractive properties as mentioned in the previous section. 

\subsection{MOO solvers for MTL problems}
\begin{figure} %
    \centering 
    \subfloat[Agg-COSMOS]{\includegraphics[width=\pentawidth \textwidth]{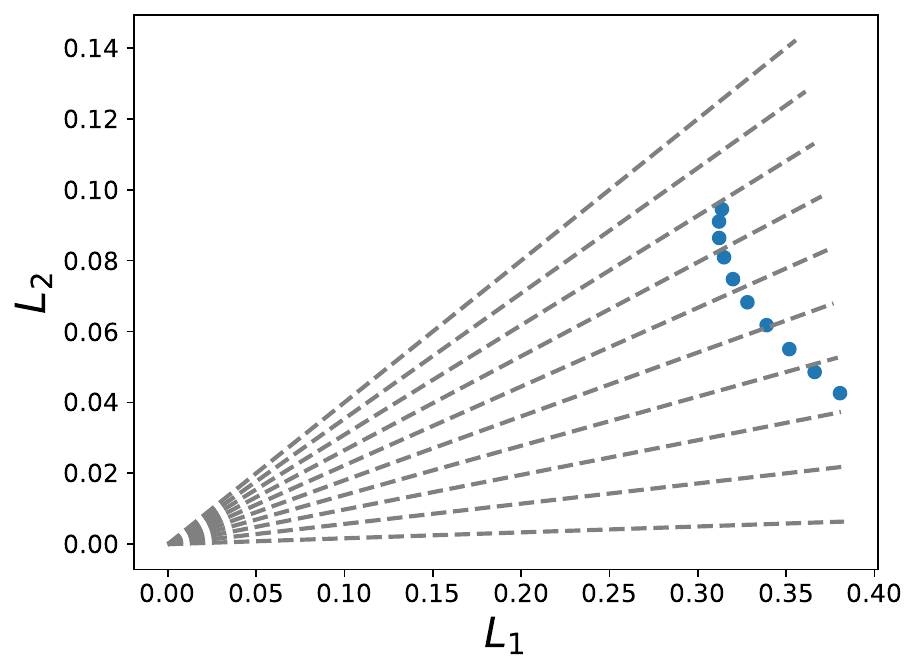}}
    \subfloat[Agg-LS]{\includegraphics[width= \pentawidth \textwidth]{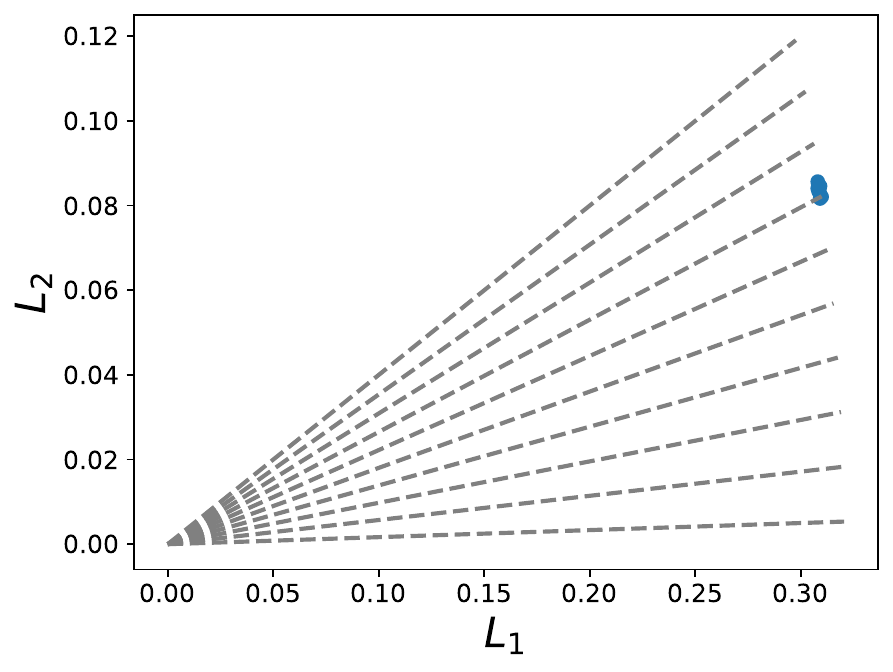}} 
    \subfloat[Agg-PBI]{\includegraphics[width= \pentawidth \textwidth]{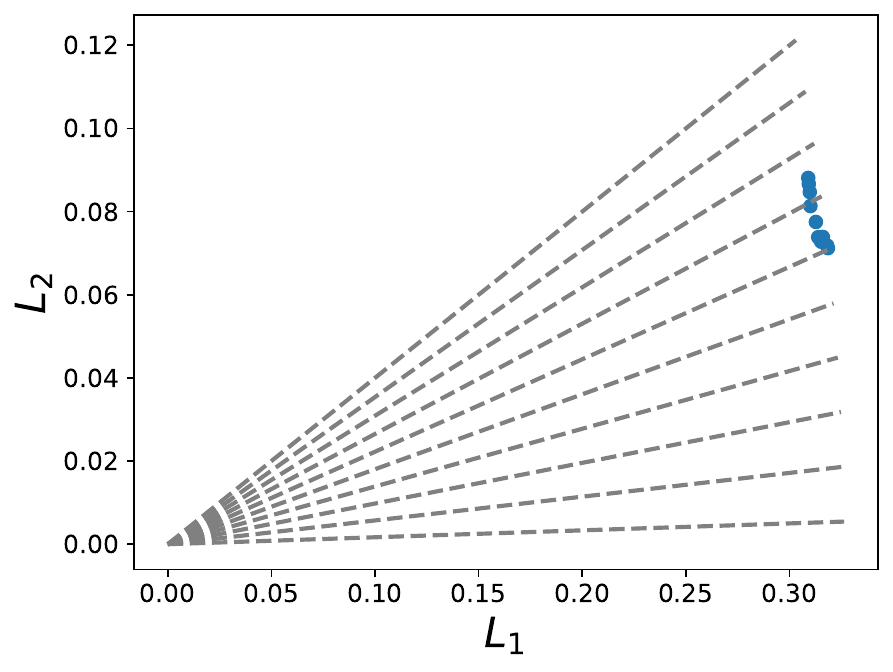}}
    \subfloat[SmoothmTche]{\includegraphics[width= \pentawidth  \textwidth]{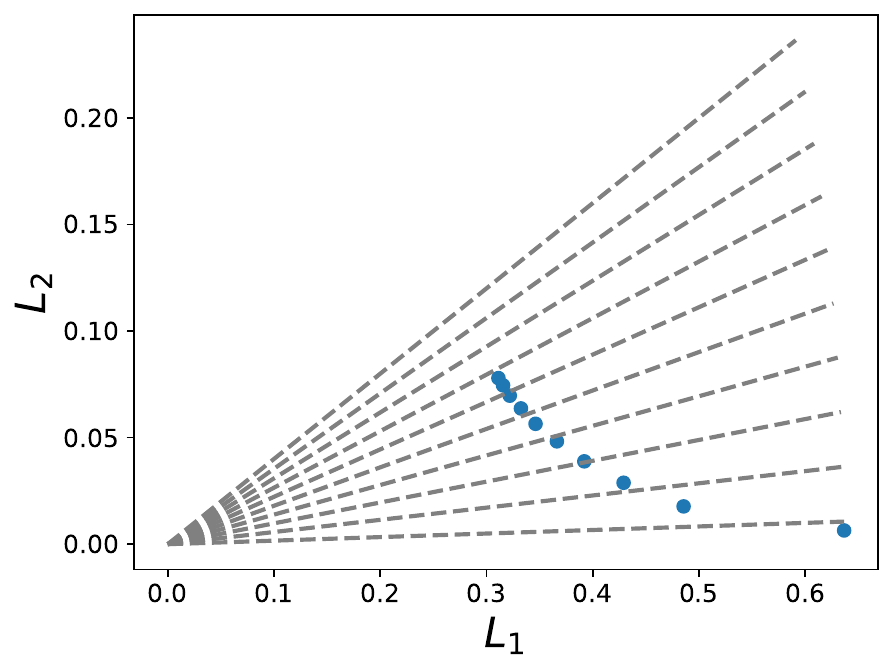}}
    \subfloat[Agg-mTche]{\includegraphics[width= \pentawidth  \textwidth]{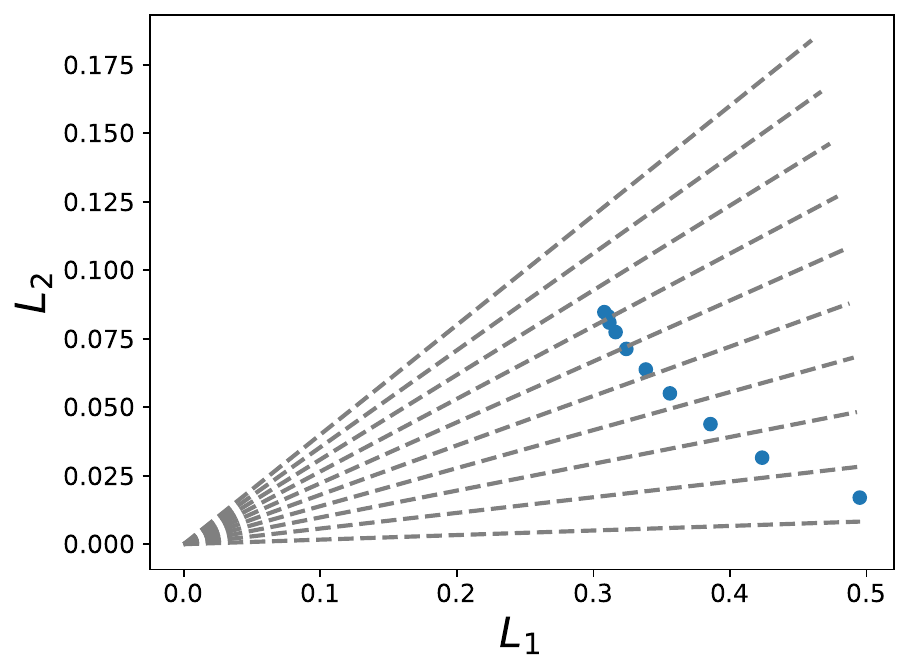}}
    \\
    \vspace{-10pt}
    \subfloat[EPO]{\includegraphics[width= \pentawidth \textwidth]{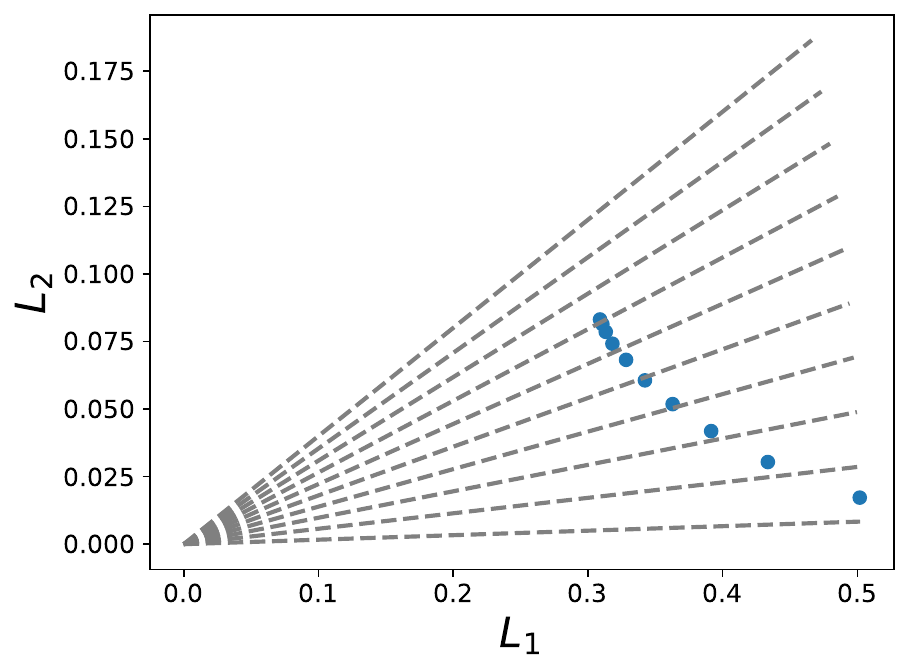}} 
    \subfloat[HVGrad]{\includegraphics[width= \pentawidth \textwidth]{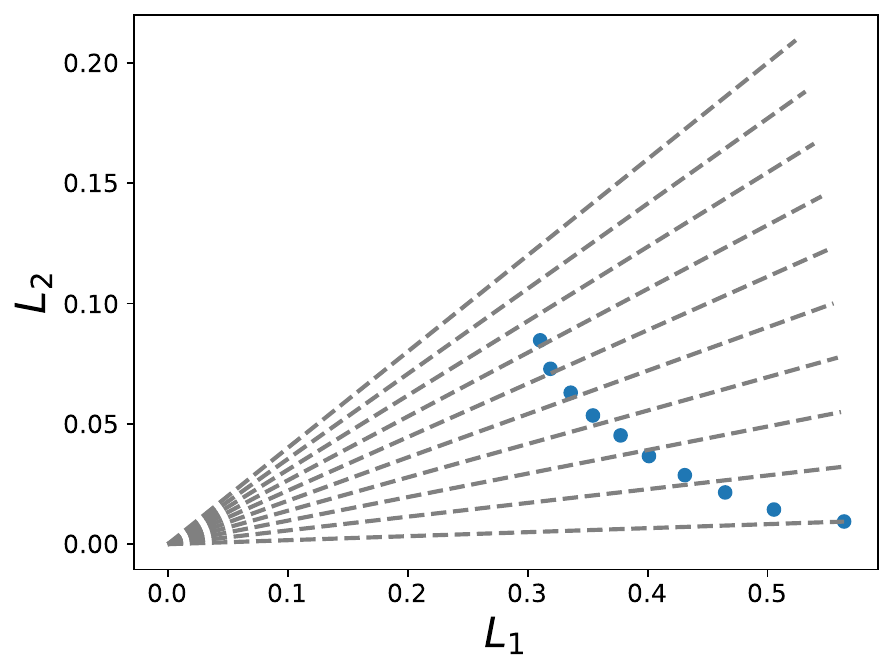}}
    \subfloat[MGDA-UB]{\includegraphics[width= \pentawidth \textwidth]{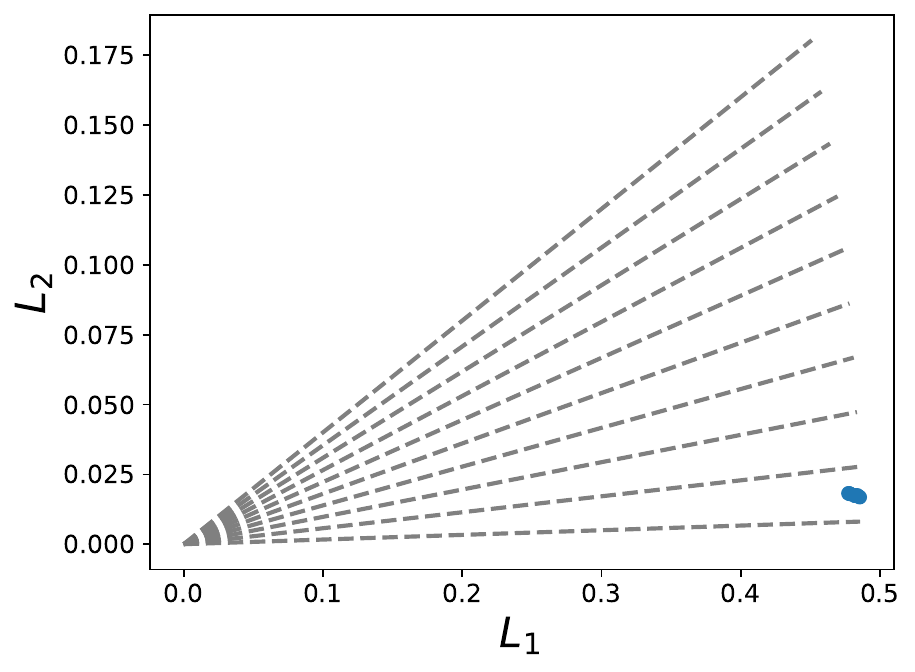}}
    \subfloat[MOO-SVGD]{\includegraphics[width= \pentawidth \textwidth]{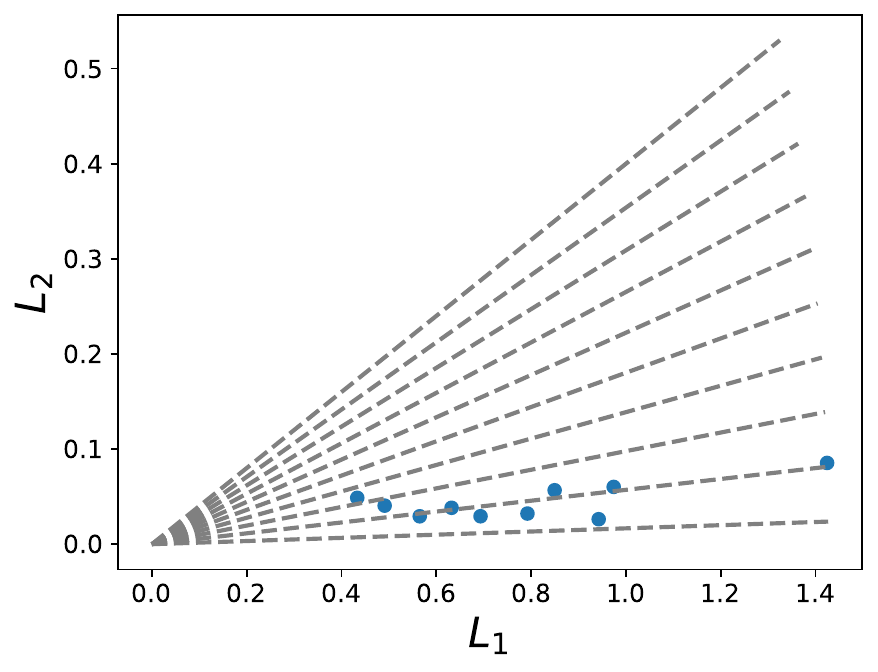}}
    \subfloat[PMTL]{\includegraphics[width= \pentawidth \textwidth]{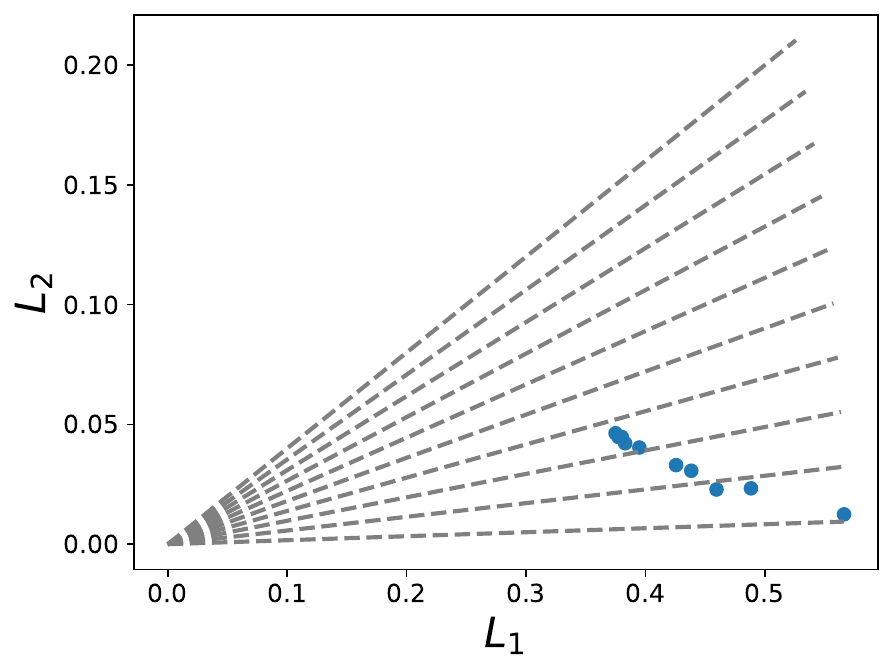}}
    \\
    \caption{Finite Pareto solutions by different solvers on Adult problem.} \label{fig:finit_mtl}
\end{figure}

We evaluate the performance of finite Pareto solvers on the Adult dataset, a multitask fairness classification problem. The decision variable $\vtheta$ represents the parameters of a fully-connected neural network with $|\vtheta|=28033$. The first objective is cross-entropy loss, and the second is the DEO loss~\cite{ruchte2021scalable}[Eq. 6]. Key findings from the \Cref{fig:finit_mtl,tab:finite_mtl} are as follows:

\textcircled{1}. Agg-LS has two drawbacks: (1) it cannot identify the non-convex part of a PF (as previous section mentioned), and (2) the relationship between preference vectors and Pareto objectives is unknown; different preference vectors may yield duplicate solutions. Agg-PBI and Agg-COSMOS only find a small portion of the PF. 
\textcircled{2}. Agg-Smooth mTche and Agg-mTche perform well on this task, as they can find (approximate) `exact' Pareto solutions. Once the Pareto front range is known, diverse solutions can be easily found using uniform preference vectors. The Random and MGDA-UB methods only find a single Pareto solution, since the position of this solution cannot be controlled by these methods. 
\textcircled{3}. Among the three methods that directly find a set of Pareto solutions (MOO-SVGD, PMTL, and HV-Grad), HV-Grad produces the most diverse solutions with the largest hypervolume. PMTL, being a two-stage method, may fail when solutions fall outside the sector due to stochastic factors. MOO-SVGD optimizes both convergence and diversity but is generally unstable based on our tests.

\textbf{Conclusion.} For convex Pareto fronts in MTL problems, \emph{Agg-LS} is recommended for computational efficiency. However, \emph{PMGDA} or \emph{EPO} may offer better convergence and preference-solution correspondence.

\begin{table}[]
\centering
\setlength\tabcolsep{2.0 pt}
\begin{threeparttable}
\caption{Numerical results of finite Pareto solutions for the Adult problem.}
\label{tab:finite_mtl}
\centering
\tiny
\begin{tabular}{l|l|l|l|l|l|l|l|l|l}
\toprule
Method                         & $l_\mathrm{min}$ & $sl_\mathrm{min}$          & Spacing           & Sparsity          & HV                & IP                & Cross Angle       & PBI               & Span              \\ 
\midrule
Agg-COSMOS                    & 0.004 (0.000)     & -0.194 (0.000)     & 0.463 (0.004)     & 0.014 (0.001)     & 0.657 (0.000)     & 0.295 (0.000)     & 2.787 (0.013)     & 0.426 (0.000)     & 0.052 (0.000)     \\ 
Agg-LS                        & 0.000 (0.000)     & -0.223 (0.000)     & \textbf{0.016 (0.005)} & 0.000 (0.000)     & 0.636 (0.001)     & \textbf{0.272 (0.001)} & 6.595 (0.028)     & 0.500 (0.001)     & 0.002 (0.000)     \\ 
Agg-PBI                       & 0.001 (0.000)     & -0.216 (0.000)     & 0.107 (0.008)     & 0.001 (0.000)     & 0.642 (0.000)     & 0.277 (0.001)     & 4.995 (0.007)     & 0.462 (0.001)     & 0.010 (0.000)     \\ 
Agg-SmoothmTche                 & 0.005 (0.001)     & -0.163 (0.000)     & 4.237 (0.029)     & 0.329 (0.003)     & 0.675 (0.001)     & 0.347 (0.001)     & 3.385 (0.053)     & 0.500 (0.002)     & 0.072 (0.000)     \\ 
Agg-mTche                     & 0.002 (0.000)     & -0.177 (0.000)     & 2.034 (0.072)     & 0.091 (0.004)     & 0.674 (0.000)     & 0.316 (0.000)     & \textbf{1.962 (0.045)} & \textbf{0.422 (0.001)} & 0.067 (0.001)     \\ 
EPO                          & 0.002 (0.000)     & -0.175 (0.000)     & 2.136 (0.054)     & 0.101 (0.004)     & 0.674 (0.001)     & 0.320 (0.000)     & 2.002 (0.025)     & 0.426 (0.001)     & 0.066 (0.000)     \\ 
MGDA-UB                      & 0.000 (0.000)     & -0.222 (0.001)     & 0.050 (0.039)     & 0.000 (0.000)     & 0.510 (0.003)     & 0.410 (0.004)     & 9.586 (0.072)     & 0.878 (0.011)     & 0.001 (0.000)     \\ 
Random                       & 0.000 (0.000)     & -0.224 (0.000)     & 0.040 (0.016)     & \textbf{0.000 (0.000)} & 0.633 (0.001)     & 0.279 (0.000)     & 5.863 (0.010)     & 0.491 (0.000)     & 0.002 (0.000)     \\ 
PMTL                         & 0.002 (0.001)     & -0.176 (0.002)     & 2.236 (0.045)     & 0.156 (0.028)     & 0.617 (0.002)     & 0.372 (0.004)     & 7.039 (0.242)     & 0.675 (0.016)     & 0.035 (0.001)     \\ 
MOO-SVGD                     & \textbf{0.049 (0.007)} & \textbf{-0.079 (0.003)} & 5.382 (4.539)     & 2.660 (1.857)     & 0.548 (0.005)     & 0.657 (0.013)     & 8.354 (0.211)     & 1.274 (0.029)     & 0.065 (0.010)     \\ 
HVGrad                       & 0.014 (0.002)     & -0.153 (0.002)     & 1.347 (0.041)     & 0.110 (0.003)     & \textbf{0.678 (0.001)} & 0.347 (0.004)     & 7.043 (0.260)     & 0.663 (0.017)     & \textbf{0.075 (0.001)} \\
\bottomrule
\end{tabular}
\begin{tablenotes}
    \small
    \item All methods were run five times with random seeds; results are presented in the form of (mean)(std).
\end{tablenotes}
\end{threeparttable}
\end{table}

\subsection{Pareto set learning on MTL}
\begin{figure}%
    \centering
    \subfloat[Agg-COSMOS]{\includegraphics[width= \pentawidth \textwidth]{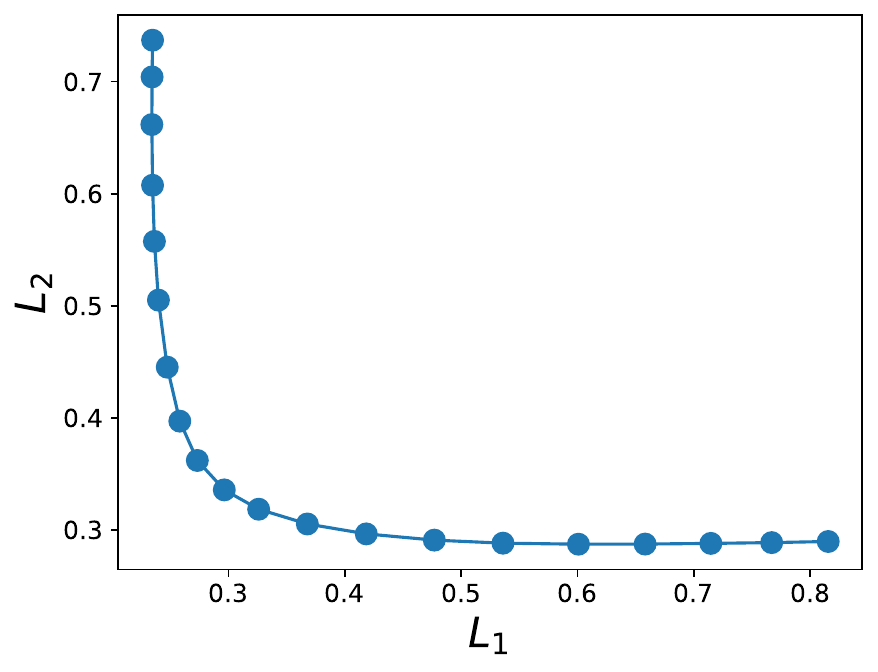}}
    \subfloat[Agg-LS]{\includegraphics[width= \pentawidth \textwidth]{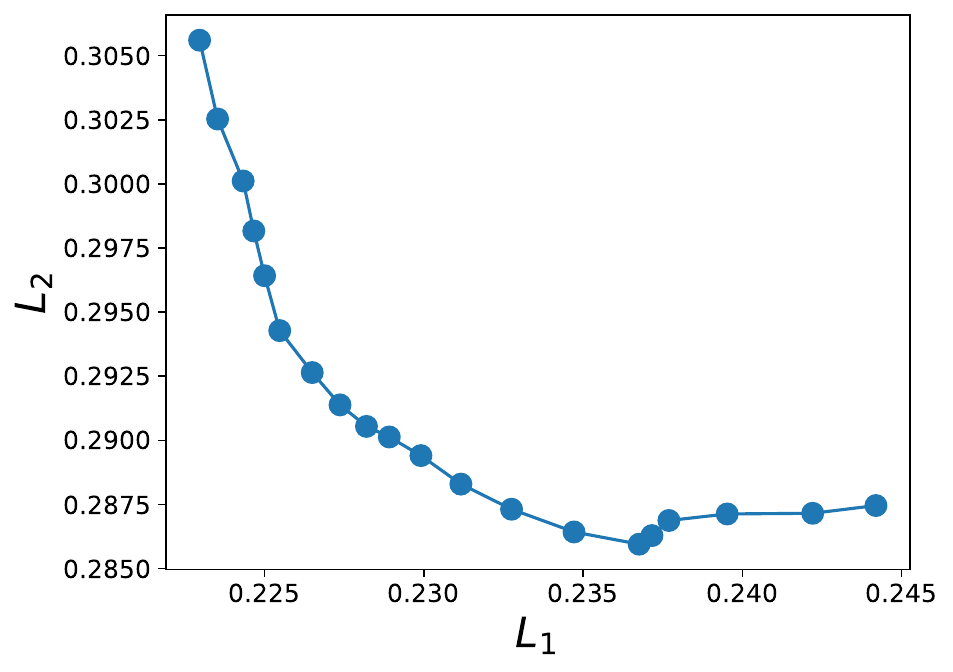}} 
    \subfloat[Agg-SmoothTche]{\includegraphics[width= \pentawidth \textwidth]{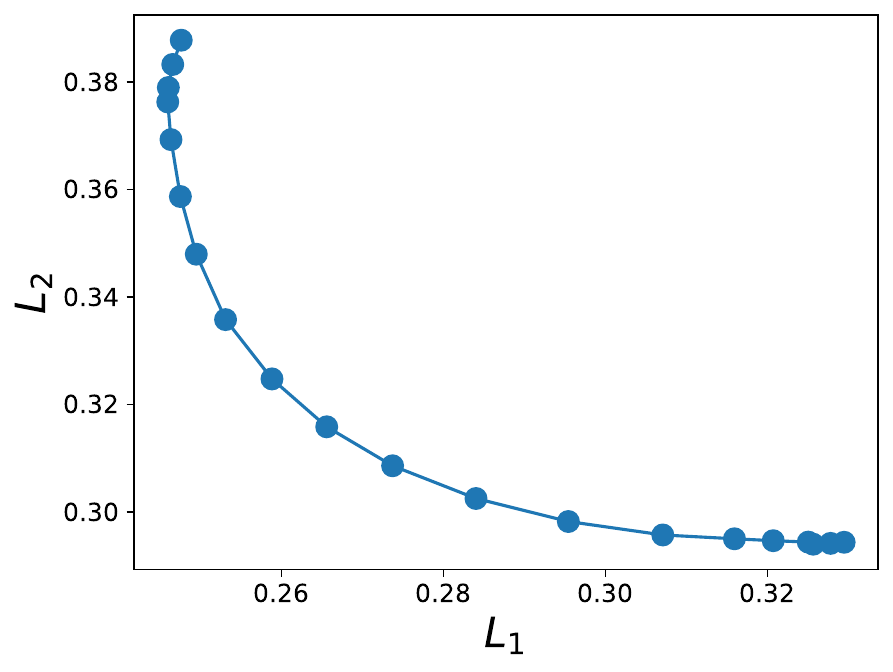}}
    \subfloat[Agg-Tche]{\includegraphics[width= \pentawidth \textwidth] {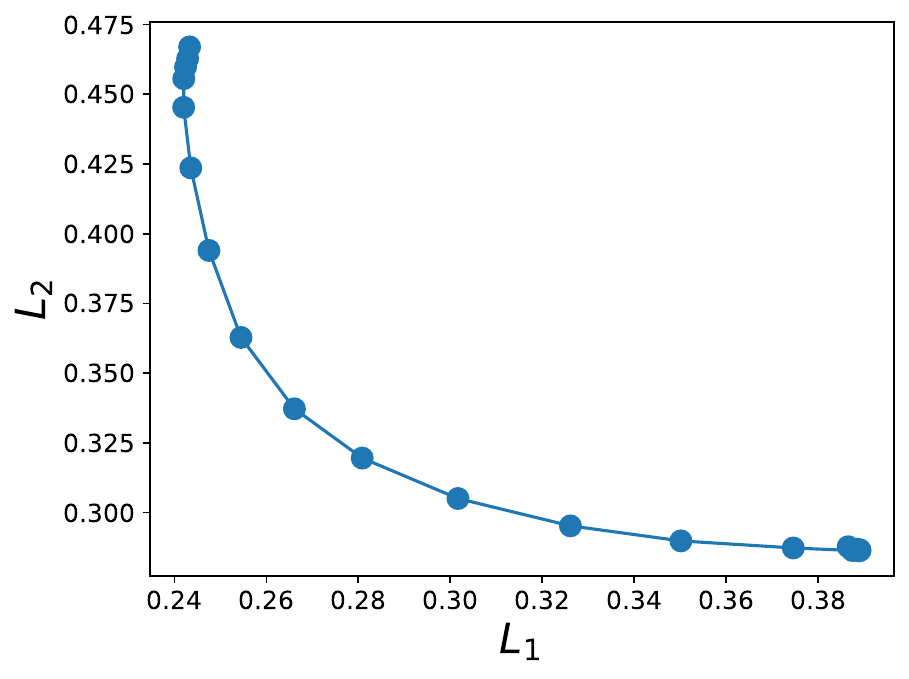}}     \\
    \vspace{-10pt}
    \subfloat[Agg-COSMOS]{\includegraphics[width= \pentawidth \textwidth]{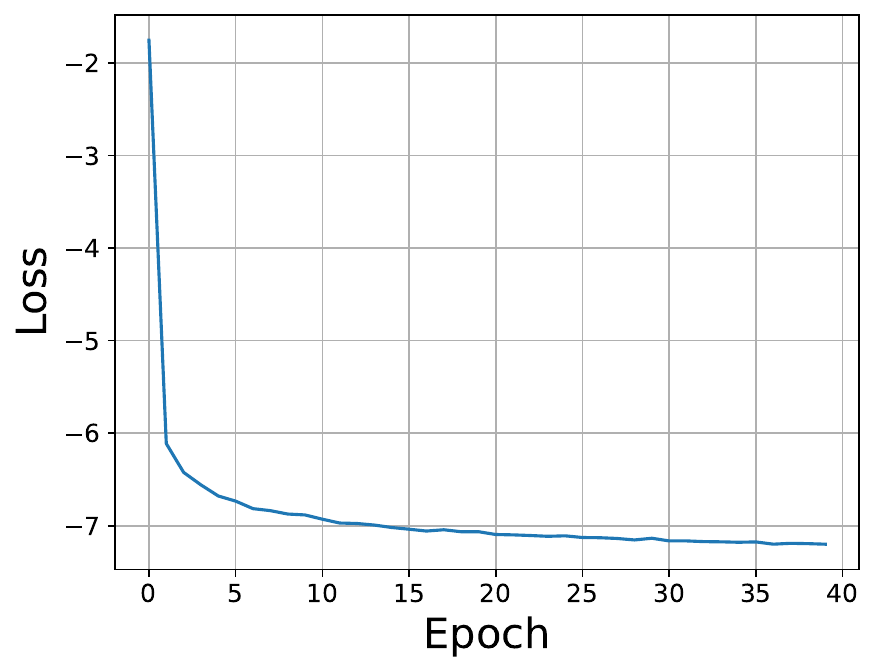}}
    \subfloat[Agg-LS]{\includegraphics[width= \pentawidth \textwidth]{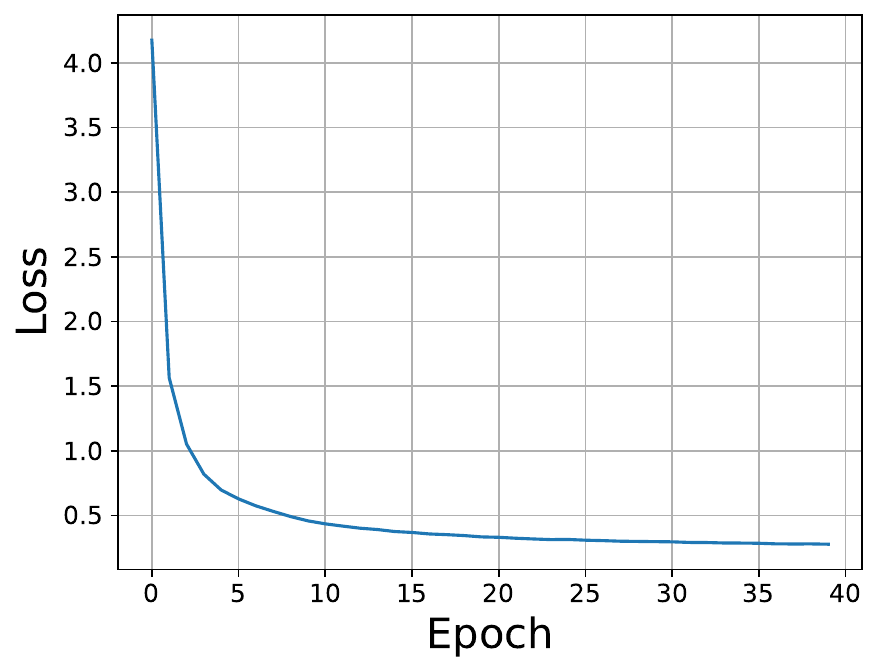}} 
    \subfloat[Agg-SmoothTche]{\includegraphics[width= \pentawidth \textwidth]{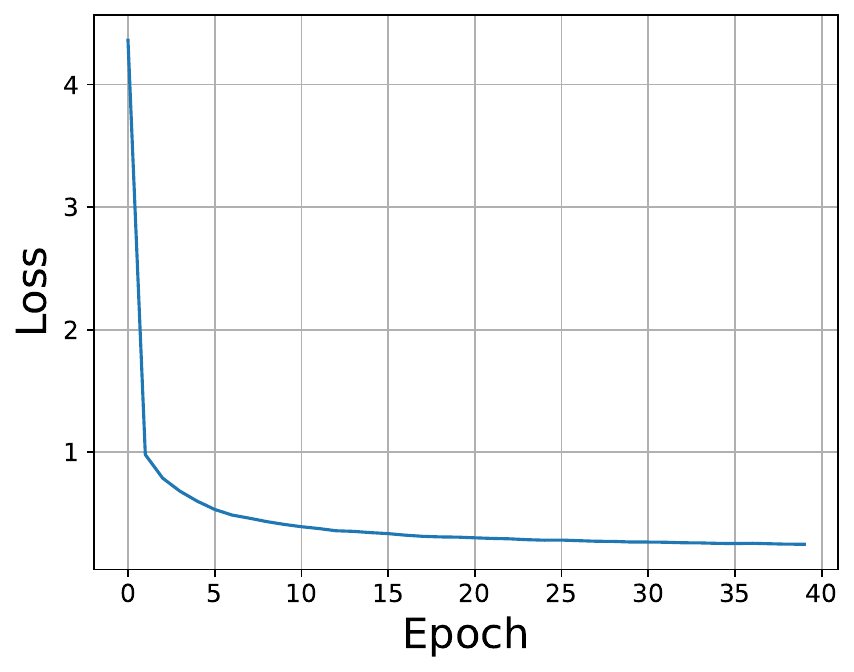}}
    \subfloat[Agg-Tche]{\includegraphics[width= \pentawidth \textwidth] {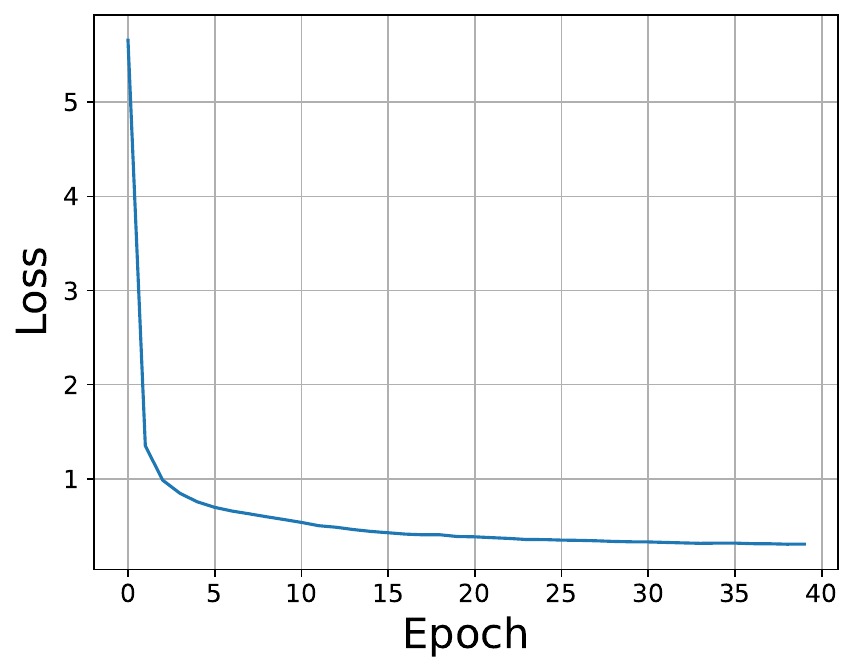}} 
    \caption{Training process for generating predicted Pareto solutions using different PSL solvers on MO-MNIST problem.} \label{fig:psl_mtl}
\end{figure}

This section presents the Pareto set learning results on the MO-MNIST problem using a hypernet architecture. The Pareto set model was trained for 20 epochs, optimizing approximately 3.24M parameters, with the first and second objectives being the cross-entropy losses for the top-right and bottom-left images. EPO-based or PMGDA-based PSL is not very suitable for this task since manipulating gradient of 3.2M parameters is not efficient. 
Empirically, for regression tasks, the PF shape is nearly convex. Therefore, Agg-LS is adequate to recover the whole PF. From \Cref{fig:psl_mtl,tab:psl_mtl}, we have that Agg-LS significantly outperforms other methods, evidenced by the HV of Agg-LS is much larger than other methods. Furthermore, the training losses across all methods are almost stable after 40 epochs. From the figure, to further reduce the training loss, it needs much more computational resources. \\
\textbf{Conclusion.} For MTL problems with a convex PF, it is highly recommended to use \emph{Agg-LS-based} PSL for the sake of computational efficiency.

\begin{table}[]
\tiny
\caption{Pareto set learning results on MO-MNITS problem.}
\label{tab:psl_mtl}

\setlength\tabcolsep{2.0 pt}
\centering
\begin{tabular}{l|l|l|l|l|l|l|l|l|l}
\toprule
Method & $l_\mathrm{min}$ & $sl_\mathrm{min}$ & Spacing & Sparsity & HV                & IP & Cross Angle & PBI & Span \\
\midrule
COSMOS                     & \textbf{0.033 (0.002)}     & \textbf{-0.152 (0.007)}     & 0.967 (0.202)              & 0.278 (0.038)              & 0.512 (0.015)              & 0.535 (0.028)              & \textbf{9.208 (0.541)}     & 1.221 (0.050)              & \textbf{0.497 (0.043)}     \\ 
Agg-LS                         & 0.001 (0.000)              & -0.286 (0.002)              & 0.068 (0.029)              & 0.000 (0.000)              & \textbf{0.557 (0.004)}     & \textbf{0.257 (0.004)}     & 27.536 (0.116)             & \textbf{1.149 (0.021)}     & 0.016 (0.003)              \\ 
Agg-PBI                        & 0.000 (0.000)              & -0.295 (0.000)              & \textbf{0.019 (0.006)}     & \textbf{0.000 (0.000)}     & 0.536 (0.008)              & 0.269 (0.005)              & 26.270 (0.117)             & 1.150 (0.019)              & 0.002 (0.001)              \\ 
SmoothTche                   & 0.001 (0.000)              & -0.248 (0.002)              & 0.440 (0.040)              & 0.012 (0.001)              & 0.538 (0.008)              & 0.281 (0.005)              & 32.244 (0.417)             & 1.471 (0.007)              & 0.087 (0.012)              \\ 
Agg-Tche                       & 0.000 (0.000)              & -0.235 (0.004)              & 0.988 (0.110)              & 0.045 (0.015)              & 0.533 (0.010)              & 0.292 (0.008)              & 35.292 (0.742)             & 1.693 (0.075)              & 0.131 (0.014) \\
\bottomrule
\end{tabular}
\end{table}

\subsection{MOBO for synthetic and real-world problems}
In this section, we test three MOBO algorithms in \algoname~on three benchmark problems, including ZDT1, RE21, VLMOP1 and VLMOP2. To ensure a fair comparison, we generate $11n-1$ initial samples using Latin Hypercube Sampling for each method. The maximum number of function evaluations is set as 200.
Our experimental results, illustrated in Figure~\ref{fig:psl_synthetic}, clearly demonstrate the rapid convergence capabilities of all three methods. DirHV-EGO, PSL-DirHV-EI, and PSL-MOBO not only efficiently navigate the solution space but also quickly reach optimal solutions. This highlights the robustness and effectiveness of our implemented algorithms in handling different types of MOPs.

\begin{figure}%
    \centering
    \subfloat[ZDT1]{\includegraphics[width= \quadwid \textwidth]{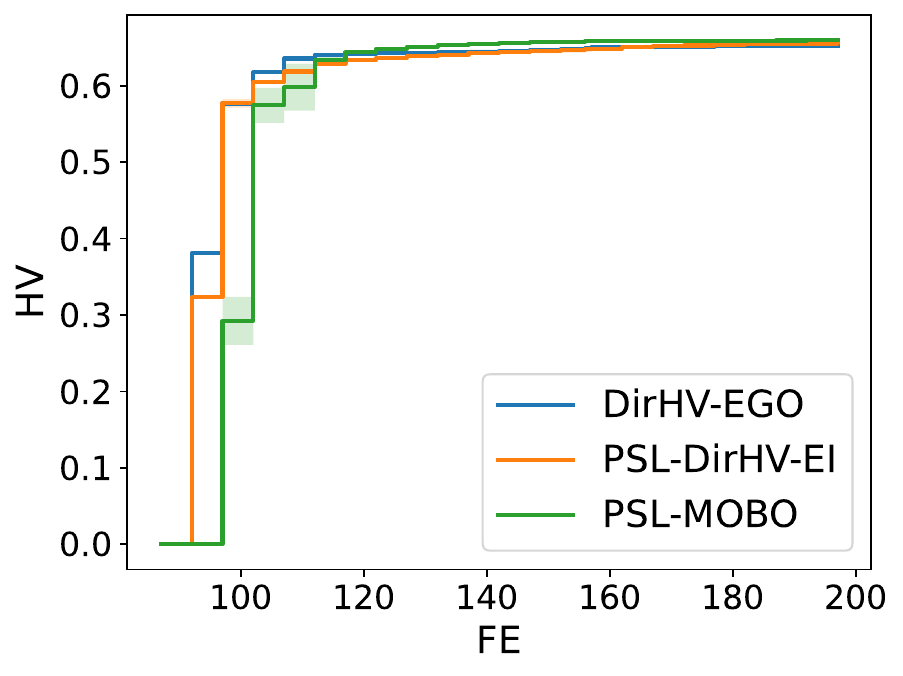}}
    \subfloat[RE21]{\includegraphics[width= \quadwid \textwidth]{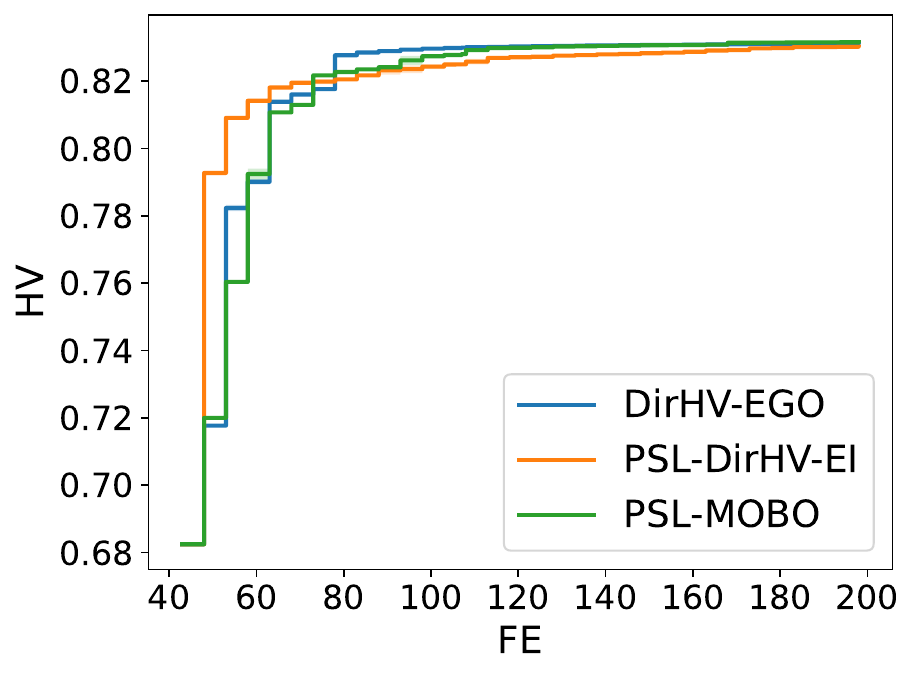}} 
    \subfloat[VLMOP1]{\includegraphics[width= \quadwid \textwidth]{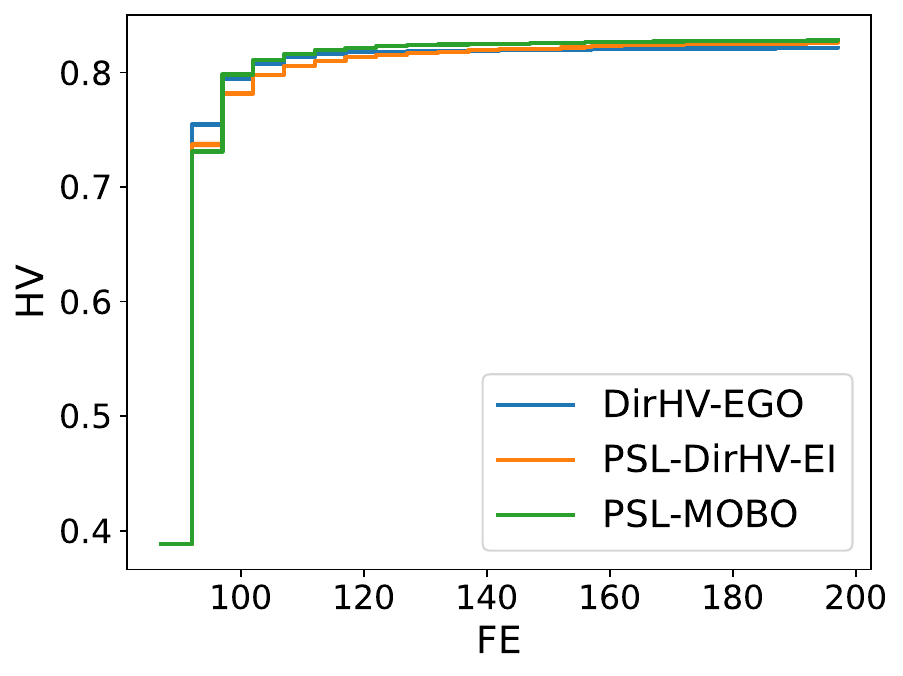}}
    \subfloat[VLMOP2]{\includegraphics[width= \quadwid \textwidth]{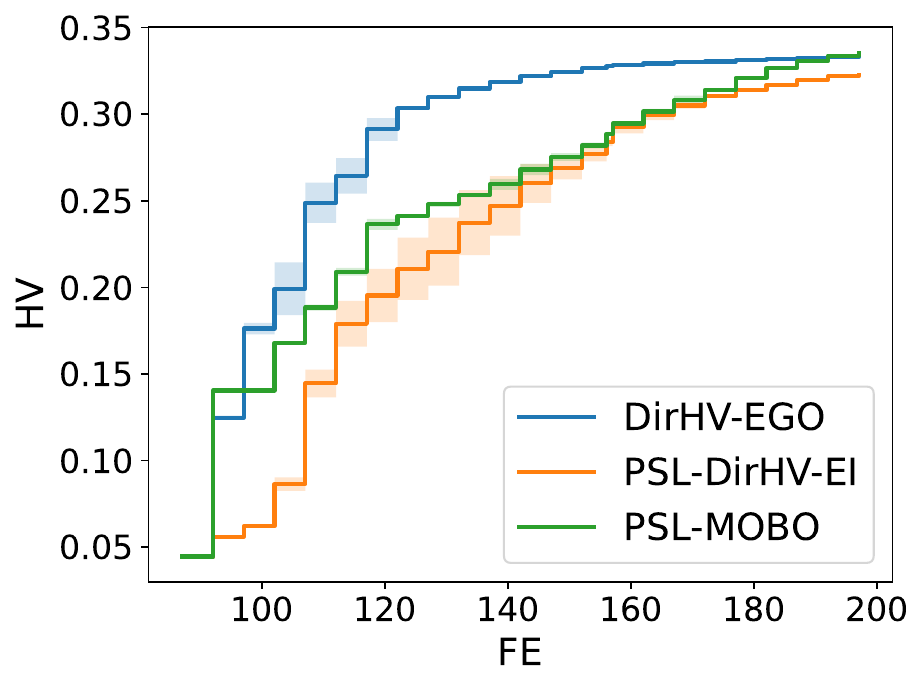}}
    \\
    \caption{HV curves on MOBO problem. Results are averaged on five random seeds. Reference point to calculate HV : [1.2, 1.2].} \label{fig:mobo}
\end{figure}

\section{Conclusion, limitations, and further works} \label{sec:conclusion}
\paragraph{Conclusion.}
We introduced the \textit{first} modern gradient-based MOO framework called \algoname~in PyTorch for the research community's convenience. 
\algoname~supports more than 20 mainstream gradient-based MOO methods;
the modular design of \algoname~further allows the library to address various MOPs via various methods in a plug-and-play manner.
\algoname~can thus be leveraged to quickly yet robustly test new MOO ideas.

\paragraph{Limitations} include: (1) rapid developments of gradient-based MOO methods makes it hard to incorporate all methods, so some effective methods may be missing; (2) gradient-based solvers may fail for problems with a number of local optimas. 

\paragraph{Future Work} includes (1) maintaining a user and development community to address issues and (2) adding newly published methods as quickly as possible.

\clearpage
\section*{Acknowledge}
\algoname~was made possible with the valuable feedback and comments from a lot of researchers. We appreciate the help from Hongzong Li (for some advice on skipping local optima), Xuehai Pan (for some software design advice), Weiyu Chen (for his code on LoRA PSL), Baijiong Lin (for his advice on automatic code testing), Prof. Hao Wang (for early email communication of HVGrad), Prof. Jingda Deng (for his advice on high-D HV calculation), Prof. Ke Shang (HV computation and RE test suit), Prof. Genghui Li, Prof. Zhichao Lu and Prof. Zhenkun Wang.

The work described in this paper was supported by the Research Grants Council of the Hong Kong Special Administrative Region, China [GRF Project No. CityU 11215622]. 

\appendix

\section{Appendix} 

\subsection{Full name and notation tables} \label{sec:fullname}
This section lists the full names of optimization methods and terms for clarity (\Cref{tab:full:name}) and provides notation in \Cref{tab:notation}. 

\subsection{Aggregation functions} \label{sec:agg}
Aggregation function convert an MOP into a single-objective optimization problem under a specific preference vector $\vlam$. Some popular aggregation functions are:

\begin{enumerate}
    \item \textbf{COSMOS}:
    \begin{equation}
        g^\mathrm{cosmos}_\vlam(\vtheta) = \vlam^\top \vL(\vtheta) - \mu \frac{\vlam^\top \vL(\vtheta)}{\|\vlam\| \|\vL(\vtheta)\|},
    \end{equation}
    where $\mu$ is a positive weight factor to align the objective vector $\vL(\vtheta)$ with the preference vector $\vlam$.
    
    \item \textbf{Linear scalarization (LS)}:
    \begin{equation}
        g^\mathrm{LS}_\vlam(\vtheta) = \sum_{i=1}^m \lambda_i L_i(\vtheta).
    \end{equation}

    \item \textbf{Tchebycheff (Tche)}:
    \begin{equation}
    g^\mathrm{Tche}_\vlam(\vtheta) = \max_{1 \leq i \leq m} \lbr{ \lambda_i (L_i(\vtheta) - z_i) },
    \end{equation}
    where $\vz$ is a reference point, usually set as the nadir point the minimal value in each objectives. 

    \item \textbf{Modified Tchebycheff (mTche)}:
    \begin{equation} \label{eqn:app:mtche}
        g^\mathrm{mTche}_\vlam(\vtheta) = \max_{1 \leq i \leq m} \left \{ \frac{L_i(\vtheta) - z_i}{\lambda_i} \right \},
    \end{equation}
    where $\vz$ is a reference point, the same as the point used in the Tchebycheff scalarization function. 
    \item \textbf{Smooth Tchebycheff (STche)}:
    \begin{equation}
        g^\mathrm{STche}_\vlam(\vtheta) = \frac{1}{h} \log \sbr{\sum_{i=1}^m \exp(h \cdot \lambda_i (L_i(\vtheta) - z_i))}.
    \end{equation}
    The Smooth Tchebycheff function uses a relaxed Smooth \(\max\) operator. The advantage of this approach is that \(g^\mathrm{STche}_\vlam(\vtheta)\) becomes a Smooth function if each objective function \(L_i(\vtheta)\) is Smooth, unlike the non-Smooth \(g^\mathrm{Tche}_\vlam(\vtheta)\). Smooth functions generally have faster convergence compared to non-Smooth ones. Similarly, we can define the Smooth modified Tchebycheff function. 

    \item \textbf{Penalty-Based Intersection (PBI):}
    \begin{equation}
        g^\mathrm{PBI}_\vlam(\vtheta) = 
        \underbrace{\frac{1}{\norm{\vlam}} \cdot \sum_{i=1}^m \lambda_i L_i(\vtheta)}_{d_1}
         + \mu \underbrace{\norm{\vL(\vtheta) - \frac{d_1}{\norm{\vlam}} \cdot \vlam}}_{d_2},
    \end{equation}
    where $\mu$ is a positive weight factor that encourage a objective to align with a preference vector $\vlam$.

    \item \textbf{$p$-norm}: 
    \begin{equation}
        g^\mathrm{pnorm}_\vlam(\vtheta) =
        {\norm{\vlam \odot \vL(\vtheta) - \vz}_p}.
    \end{equation}
    The symbol $\odot$ denotes the element-wise product between two vectors.
    \item \textbf{Augmented Achievement Scalarization Function (AASF)}:
    \begin{equation}
        g^\mathrm{AASF}_\vlam(\vtheta) = g^\mathrm{mTche}_\vlam(\vtheta) + \rho g^\mathrm{LS}_\vlam(\vtheta),
    \end{equation}
    where $\rho$ is small positive coefficient, which is usually set as 0.1. Contour curves for this functions for a bi-objective case can be found in the \href{https://libmoondocs.readthedocs.io/en/latest/gallery/aggfuns.html}{LibMOON Doc}~\footnote{\url{https://libmoondocs.readthedocs.io/en/latest/gallery/aggfuns.html}}. 
\end{enumerate}

\begin{table}[]
\caption{Short name to full name table} \label{tab:full:name}
\centering
    \begin{tabular}{ll}
    \toprule
    Short Name & Full name \\
    \midrule
    MOP & \underline{M}ultiobjective \underline{O}ptimization \underline{P}roblem\\
    SOP & \underline{S}ingleobjective \underline{O}ptimization \underline{P}roblem\\
    MOO & \underline{M}ulti\underline{O}bjective \underline{O}ptimization \\
    MOEA & \underline{M}ulti\underline{O}bjective \underline{E}volutionary \underline{A}lgorithm \\
    MOBO & \underline{M}ulti\underline{O}bjective \underline{B}aysian \underline{O}ptimization \\
    PSL & \underline{P}areto \underline{S}et \underline{L}earning \\
    PS & \underline{P}areto \underline{S}et \\ 
    PF & \underline{P}areto \underline{F}ront \\ 
    `exact' Pareto solution & The corresponding Pareto objective aligns with a given preference vector \\
    \midrule
    ES & Evolutionary strategy \\ 
    BP & Backward propagation \\
    \midrule
    PMTL \cite{lin2019pareto} & \underline{P}areto \underline{M}ulta-\underline{T}ask \underline{L}earning \\
    MOO-SVGD \cite{liu2021profiling}  & \underline{M}ulti\underline{O}bjective \underline{O}ptimization \underline{S}tein \underline{V}ariational \underline{G}radient \underline{D}escent \\
    EPO \cite{mahapatra2020multi} & \underline{E}xact \underline{P}areto \underline{O}ptimization \\
    PMGDA \cite{zhang2024pmgda} & \underline{P}reference based \underline{M}ultiple \underline{G}radient \underline{D}escent \underline{A}lgorithm \\
    Agg-LS \cite{miettinen1999nonlinear} & \underline{Agg}regation function with \underline{L}inear \underline{S}calarization\\
    Agg-PBI \cite{zhang2007moea} & \underline{Agg}regation function with \underline{P}enalty \underline{B}ased \underline{I}ntersection \\
    Agg-Tche \cite{zhang2007moea} & \underline{Agg}regation function with \underline{Tche}bycheff scalarization \\
    Agg-mTche \cite{ma2017tchebycheff} & \underline{Agg}regation function with \underline{m}odified \underline{Tche}bycheff scalarization \\
    Agg-COSMOS \cite{ruchte2021scalable} & \underline{Agg}regation function with \underline{COSMOS} scalarization \\
    \midrule
    RE problems & Realworld problems \\
    \bottomrule
    \end{tabular}
\end{table}

\begin{table}[]
\caption{Notations used in this paper} \label{tab:notation}
\centering
    \begin{tabular}{ll}
    \toprule
    Notation & Meaning \\
    \midrule
    $\vtheta$ & The decision variable of an MOP. \\
    $\vphi$ & The decision variable of a Pareto set model. \\
    $m$ & Number of objectives. \\
    $n$ & Number of decision variables. \\
    $K$ & Number of finite Pareto solutions. \\
    $\alpha_i$ & Coefficients of objective functions. \\
    $\vlam$ & A preference vector. \\
    \bottomrule
    \end{tabular}
\end{table}

\subsection{Metrics} \label{sec:metrics}
Metrics used in \algoname~can be categorized into two groups. The first group evaluates the quality of a \emph{set} of solutions $\sY = \{ \vy^{(1)}, \ldots, \vy^{(N)} \}$, with specific metrics such as IGD and FD relying on the known Pareto front for accuracy.
The second group of metrics assesses the quality of \emph{individual} solutions $\vy$ when a preference vector $\vlam$ is provided.

Group 1: Metrics for a set of solutions.
\begin{enumerate}
    \item Hypervolume (HV) ($\uparrow$) \cite{guerreiro2020hypervolume}: This metric evaluates both the convergence to the Pareto Front (PF) and the diversity of solutions. A low HV value indicates poor convergence, while high HV values imply better performance. The hypervolume is calculated as the volume dominated by at least one solution in the set $\sS$ with respect to a reference point $\vr$:
    \begin{align*}
        \mathrm{HV}_\vr(\sS) = \mathrm{Vol}( {\vy \; | \; \exists \vy' \in \sS, \vy' \preceq \vy \preceq \vr } ).
    \end{align*}
    \item Inverted Generational Distance (IGD) \cite{ishibuchi2015modified}: IGD measures the average distance between points in a reference set $\sZ$ and the nearest solutions in the set $\sS$: \begin{align*}
        \operatorname{IGD}(\sS) = \frac{1}{|\sZ|}\left( \sum_{i=1}^{|\sZ|} \min_{\vy' \in \sS} \rho(\vy^{(i)}, \vy')^2 \right)^{1/2}.
    \end{align*}
    \item Fill Distance (FD) \cite{zhang2024umoea}: This metric calculates the covering radius of a set of solutions $\sS$, defined as the maximum minimum distance from any point in the reference set $\sZ$ to the nearest solution in $\sZ$:
    \begin{equation}
        \mathrm{FD}(\sS) = \max_{\vy' \in \sZ} \min_{\vy \in \sS} \rho(\vy, \vy').
    \end{equation}
    \item
    Minimal Distance ($\mathrm{l}_{\min}$): This metric captures the smallest pairwise distance among all objectives:
    \begin{align*}
        \mathrm{l}_{\min} = \min_{1 \leq i < j \leq N} \rho(\vy^{(i)}, \vy^{(j)})
    \end{align*}
    where $\rho()$ denotes the Euclidean distance.
    \item Smooth Minimal Distance ($\mathrm{sl}_{\min}$): This metric is a ``smooth-min" version of the minimal distance function, defined as:
    \begin{equation}
        \mathrm{sl}_{\min} = -\frac{1}{h \cdot k(k-1)} \log \left( \sum_{1 \leq i < j \leq N} \exp\left(-h \rho\left(\vy^{(i)}, \vy^{(j)} \right)\right) \right). 
    \end{equation}

    \item Spacing: This metric measures the standard deviation of the minimal distances from one solution to others, with lower values indicating a more uniform distribution of objective vectors: \begin{equation}
        \mathrm{spacing} = \frac{1}{N} \sum_{i=1}^N (d_i - \bar{d})^2, \qquad \bar{d} = \frac{1}{N} \sum_{i=1}^N d_i, \qquad d_i = \min_{1 \leq i \neq j \leq N} \rho(\vy^{(i)}, \vy^{(j)}).
    \end{equation}
    
    \item Span: This metric evaluates the range (span) of solutions in their minimal dimension, defined as: \begin{equation}
        \mathrm{Span} = \min_{1 \leq i \leq m} \max_{1 \leq k < l \leq N} | y_i^{(k)} - y_i^{(l)}|.
    \end{equation}
\end{enumerate}

Group 2: Metrics for individual solutions.
\begin{enumerate}
    \item Penalty-based Intersection (PBI): This metric is a weighted sum of two distance functions $d_1$ and $d_2$, given by $\mathrm{PBI} = d_1 + \mu d_2$, where \begin{equation}
        d_1 = \frac{\langle \vy - \vz, \vlam \rangle}{\norm{\vlam}}, \qquad d_2 = \norm{\vy - (d_1 \vlam + \vz)}.
    \end{equation}
    \item Inner Product (IP): This metric measures the alignment of the objective vector $\vy$ with the preference vector $\vlam$:
    \begin{equation}
        \mathrm{IP} = \langle \vy, \vlam \rangle.
    \end{equation}
    \item Cross Angle ($\vartheta$): For bi-objective problems, this metric measures the angular difference between the objective vector and the preference vector:
    \begin{equation}
        \vartheta = \norm{\arctan(y_2 / y_1) - \arctan(\lambda_2 / \lambda_1)}.
    \end{equation}
\end{enumerate}

Those metrics are summarized in \Cref{tab:metrics} and also can be found in the \href{https://libmoondocs.readthedocs.io/en/latest/apis/Libmoon_metric.html}{\algoname~document}. 

\begin{table}[]
\tiny 
\centering
\caption{Supported metrics.} \label{tab:metrics}
    \begin{tabular}{lll}
    \toprule
    Metrics       & Full name & Descriptions \\
    \midrule
    HV            & Hypervolume & Hypervolume value of the dominated region. \\
    IGD           & Inverted general distance & The average general distance between a set and the true PF. \\
    FD            & Fill distance & The radius of a set to cover the true PF.  \\
    $\mathrm{l}_{\min}$  & Minimal distance & The minimal pairwise distance of a set. \\
    $\mathrm{sl}_{\min}$ & Smooth minimal distance & The smooth minimal pairwise distance of a set.                      \\
    Spacing       & - & The standard deviation of the minimal distances to other solutions. \\
    Span          & - & The range of a set. \\
    PBI           & Penalty-based intersection   & The weighted sum distance between inner product and distance to a preference vector. \\
    IP            & Inner product & The inner product between a preference and an objective vector. \\
    $\vartheta$ & Cross product & The cross angle between a preference and an objective vector.\\
    \bottomrule
    \end{tabular}
\end{table}

\subsection{GPU acceleration} \label{sec:gpu}
\begin{figure}
    \centering
    \includegraphics[width=0.4\textwidth]{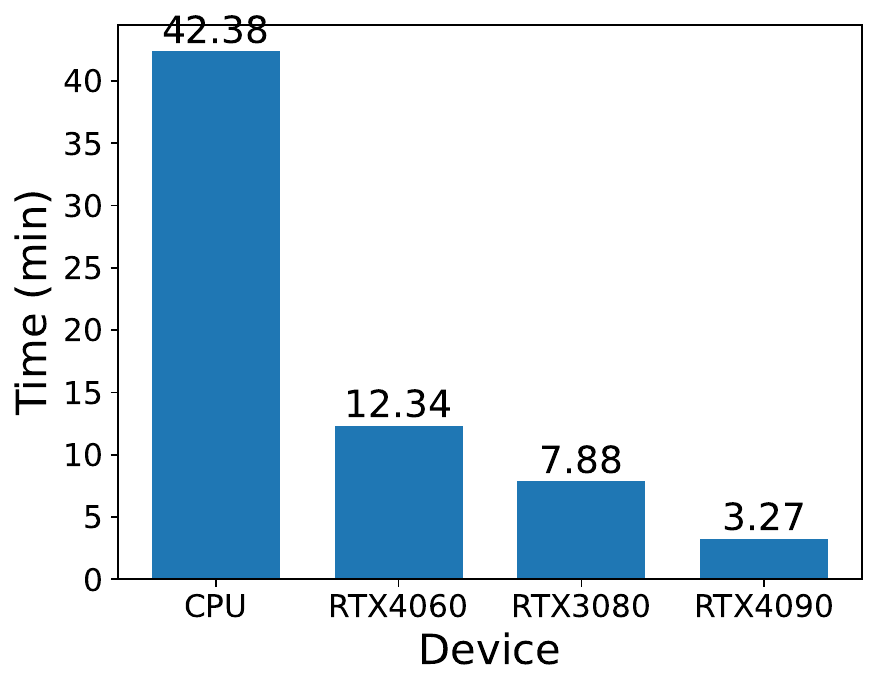}
    \caption{
    \textbf{Running time for Pareto set learning on the MO-MNIST problem using different devices using 3M parameters.}
    }
    \label{fig:running_time}
\end{figure}

We evaluate \algoname~performance on Pareto set learning for the MO-MNIST problem across various platforms (CPU, RTX 3080, 4060, 4090). Running times are detailed in \Cref{tab:running_time} and visualized in \Cref{fig:running_time}. The table and figure show a significant reduction in time (about one-third) when using a personal GPU compared to a CPU. The RTX 4090 further reduces time by approximately 25\% compared to the RTX 4060.

\begin{table}[]
\centering
\begin{threeparttable}
    \caption{Running time (minutes) for different platforms on MO-MNIST problem.}
    \label{tab:running_time}
    
    \begin{tabular}{lccc}
    \toprule
    \textbf{Platform} & \textbf{MO-MNIST} & \textbf{MO-Fashion} & \textbf{Fashion-MNIST} \\
    \midrule
    \textbf{CPU} & 43.43 & 44.45 & 46.45 \\ 
    \textbf{RTX 4060 (8G)} & 14.72 & 13.21 & 12.43 \\ 
    \textbf{RTX 3080 (10G)} & 7.88 & 7.16 & 7.17 \\ 
    \textbf{RTX 4090 (24G)} & 3.27 & 3.27 & 3.27 \\
    \bottomrule
    \end{tabular}
    \begin{tablenotes}
        \small
        \item  We run all datasets for 100 epochs using 3M parameters on a 13th Gen Intel(R) Core(TM) i9-13900HX CPU.
    \end{tablenotes}
\end{threeparttable}
\end{table}

\subsection{License, usage, and code dependence} \label{sec:license}
The license used for Adult/Compas/Credit follows Creative Commons Attribution 4.0 International (CC BY 4.0), Database Contents License (DbCL) v1.0, and CC0: Public Domain, respectively. For academic use of \algoname, please cite our paper or GitHub. Commercial use requires author permission.   

Some part codes of \algoname~follows (1) COSMOS~\footnote{\url{https://github.com/ruchtem/cosmos}}, (2) PHN~\footnote{\url{https://github.com/AvivNavon/pareto-hypernetworks}}, and (3) HVGrad~\footnote{\url{https://github.com/timodeist/multi_objective_learning}}.

\clearpage
\bibliographystyle{unsrt}  
\bibliography{moon}

\clearpage
\section*{Checklist}


\begin{enumerate}

\item For all authors...
\begin{enumerate}
  \item Do the main claims made in the abstract and introduction accurately reflect the paper's contributions and scope?
    \answerYes{}. 
  \item Did you describe the limitations of your work?
    \answerYes{}, see \Cref{sec:conclusion}. 
  \item Did you discuss any potential negative societal impacts of your work?
    \answerNA{}. \algoname~is a basic optimization library and we do not see direct societal impacts. 
  \item Have you read the ethics review guidelines and ensured that your paper conforms to them?
    \answerNA{}. 
\end{enumerate}

\item If you are including theoretical results...
\begin{enumerate}
  \item Did you state the full set of assumptions of all theoretical results?
    \answerYes{}. The assumption is in the theorem itself, e.g., in \Cref{thm:po}. 
	\item Did you include complete proofs of all theoretical results?
    \answerNo{}. This theorem is a restatement of previous results. We have give both the thereom and its proof proper citations. 
\end{enumerate}

\item If you ran experiments (e.g. for benchmarks)...
\begin{enumerate}
  \item Did you include the code, data, and instructions needed to reproduce the main experimental results (either in the supplemental material or as a URL)?
    \answerYes{}. These instruments are provided in the \algoname~Github page: \url{https://github.com/xzhang2523/libmoon}. 
  \item Did you specify all the training details (e.g., data splits, hyperparameters, how they were chosen)?
    \answerYes{}. See source code of \algoname. 
	\item Did you report error bars (e.g., with respect to the random seed after running experiments multiple times)?
    \answerYes{}. We have reported standard derivation results. 
	\item Did you include the total amount of compute and the type of resources used (e.g., type of GPUs, internal cluster, or cloud provider)?
    \answerYes{}. See first paragraph of \Cref{sec:results}. 
\end{enumerate}

\item If you are using existing assets (e.g., code, data, models) or curating/releasing new assets...
\begin{enumerate}
  \item If your work uses existing assets, did you cite the creators?
    \answerYes{}. See \Cref{sec:license}.  
  \item Did you mention the license of the assets?
    \answerYes{}. See \Cref{sec:license}. 
  \item Did you include any new assets either in the supplemental material or as a URL?
    \answerYes{}. Code are provided in \url{https://github.com/xzhang2523/libmoon}.  
  \item Did you discuss whether and how consent was obtained from people whose data you're using/curating?
    \answerYes{}. \algoname~uses public data such as multiobjective classification and fairness classification. 
  \item Did you discuss whether the data you are using/curating contains personally identifiable information or offensive content?
    \answerNA{}. 
\end{enumerate}

\item If you used crowdsourcing or conducted research with human subjects...
\begin{enumerate}
  \item Did you include the full text of instructions given to participants and screenshots, if applicable?
    \answerNA{}. 
  \item Did you describe any potential participant risks, with links to Institutional Review Board (IRB) approvals, if applicable?
    \answerNA{}
  \item Did you include the estimated hourly wage paid to participants and the total amount spent on participant compensation?
    \answerNA{}
\end{enumerate}

\end{enumerate}

\end{document}